\documentclass[useAMS,usenatbib]{mn2e}
\usepackage{graphicx}
\newcommand{\msun}{{M$_\odot$}}
\title[Simulated Circumgalactic Medium]{Constraints on Hydrodynamical Subgrid Models from Quasar Absorption Line Studies of the Simulated Circumgalactic Medium}

\author[C.B. Hummels et al.]{Cameron B. Hummels$^{1}$\thanks{E-mail:chummels@astro.columbia.edu}, Greg L. Bryan$^{1}$, Britton D. Smith$^{2}$ and
\newauthor
Matthew J. Turk$^{1}$\\
$^{1}$Department of Astronomy \& Astrophysics, Columbia University, New York, NY 10027 USA\\
$^{2}$Department of Physics and Astronomy, Michigan State University, East Lansing, MI 48824 USA}

\begin{document}

\date{}

\pagerange{\pageref{firstpage}--\pageref{lastpage}} \pubyear{2012}

\maketitle

\label{firstpage}

\begin{abstract}
Cosmological hydrodynamical simulations of galaxy evolution are increasingly able to produce realistic galaxies, but the largest hurdle remaining is in constructing subgrid models that accurately describe the behavior of stellar feedback.  As an alternate way to test and calibrate such models, we propose to focus on the circumgalactic medium.  To do so, we generate a suite of adaptive-mesh refinement (AMR) simulations for a Milky-Way-massed galaxy run to $z=0$, systematically varying the feedback implementation.  We then post-process the simulation data to compute the absorbing column density for a wide range of common atomic absorbers throughout the galactic halo, including H I, Mg II, Si II, Si III, Si IV, C IV, N V, O VI, and O VII.  The radial profiles of these atomic column densities are compared against several quasar absorption line studies, to determine if one feedback prescription is favored.  We find that although our models match some of the observations (specifically those ions with lower ionization strengths), it is particularly difficult to match O VI observations.  There is some indication that the models with increased feedback intensity are better matches.  We demonstrate that sufficient metals exist in these halos to reproduce the observed column density distribution in principle, but the simulated circumgalactic medium lacks significant multiphase substructure and is generally too hot.  Furthermore, we demonstrate the failings of inflow-only models (without energetic feedback) at populating the CGM with adequate metals to match observations even in the presence of multiphase structure.  Additionally, we briefly investigate the evolution of the CGM from $z=3$ to present.  Overall, we find that quasar absorption line observations of the gas around galaxies provide a new and important constraint on feedback models.
\end{abstract}


\section{Introduction}
\label{intro}

Following the flow of baryons in and out of galaxies is necessary to understand the star formation rate, stellar content and overall structure of galaxies of all sizes.  Significant progress has been made in our theoretical comprehension of the inflow side of this equation, with the improved understanding of cold and hot-mode accretion \citep[e.g.][]{1977ApJ...215..483B, 2003MNRAS.345..349B, Keres:2005gb, 2008MNRAS.390.1326O, Keres:2009bq, 2011MNRAS.417.2982F, 2011MNRAS.414.2458V}.  In this picture, halos with masses below about $3 \times 10^{11}$ \msun\ accrete most of their gas in the form of ("cold") $\sim 10^4$ K filamentary flows, while those that are more massive develop an accretion shock and tend to accrete gas in a smooth, hot form.  This implies that Milky Way progenitor systems, at high redshift, received baryons in the cold mode, but that present-day large spirals do not.  

Although the existence of these cold streams is robustly predicted by $\Lambda$CDM simulations, observational confirmation remains challenging.  Using high-resolution SPH simulations, \citet{2011MNRAS.412L.118F} demonstrated that the H I covering fraction of cold streams near damped Ly-alpha absorbers (DLAs) and Lyman-limit systems (LLSs) is 3\% to 10\% respectively at $z\sim2$, an observational signal easily overwhelmed by galactic winds. \citet{2011MNRAS.413L..51K} found that the predicted covering fraction of dense gas $N_{\rm HI } > 10^{20}$ cm$^{-2}$ at $z \sim 2-3$ that would give rise to observable C II absorption in background quasi-stellar objects (QSOs) is only about 5\% for Milky-Way-mass halos. Despite these low covering fractions, \citet{2011MNRAS.418.1796F} found that the cold streams at high-redshift contribute significantly to the observed population of Lyman-limit systems \citep[see also][]{2011ApJ...736...42R}, and that the transition from low to high mass should be accompanied by a disappearance of Lyman-limit absorbers at moderate to high redshift \citep{2011ApJ...735L...1S}.  At low redshift, the inflow picture is even less clear, as accretion is predicted to transition out of the cold-mode.  Detailed predictions are currently lacking, although see \citet{Joung:2012wj, 2012ApJ...749..181F} for a detailed comparison of H I gas properties between simulations and observations at $z \sim 0$.

Galactic winds are the other side of the baryon flow equation \citep{2012MNRAS.421...98D}, and evidence from the current stellar content of galaxies indicates that these must carry a substantial amount of mass and metals \citep{2006MNRAS.368..715M, Oppenheimer:2008bu, Behroozi:2010ja}.  Numerical simulations have struggled to reproduce these large mass outflows \citep{Katz:1992gu, Abadi:2003bs, 2012ApJ...749..140H}, but the addition of subgrid models can produce high mass winds \citep[e.g.][]{Springel:2003eg, Stinson:2006id, Oppenheimer:2006eq, Scannapieco:2006ij, 2012MNRAS.422.1231G}.  However, a detailed understanding of what drives winds is still lacking, with even the basic physical mechanism (e.g. ram pressure from supernova bubbles or radiation pressure) in dispute \citep[e.g.][]{2011ApJ...735...66M, 2012ApJ...750...55S, 2012MNRAS.421.3522H}.

One important way to probe both the inflow and outflow of baryon, and an important test for any cosmological galaxy formation model, is the observed correlation between absorption lines in background quasars and the projected distance (and velocity offset) to the associated galaxy.  A substantial body of such data has been built up at high redshift \citep{2004ApJ...606...92S, 2010ApJ...717..289S}.  These data are beginning to be compared against simulations at high redshift \citep[e.g.][]{2011ApJ...737L..37A, 2011MNRAS.411..826T, 2011MNRAS.418.1796F, 2011MNRAS.416.1723B, Goerdt:2012vk, Shen:2012ux}.  At low redshift, where we focus, there exists a large body of data combining galaxy imaging and QSO spectroscopy \citep{Chen:2001hj, Chen:2008cd, 2008ApJ...683...22T, 2008ApJ...672L..21Y, Chen:2010ir, 2011ApJ...740...91P, 2011ApJ...736....1T, Tumlinson:2011wm}, but only a few predictions from numerical simulations \citep{Stinson:2011tp, 2012ApJ...749..181F}.

In this paper, we use high-resolution cosmological simulations of a galaxy run to $z = 0$ to make predictions for a wide variety of absorption lines as a function of impact parameter and compare them against observational samples.  We include tracers of cold gas ($T \sim 10^4 - 5 \times 10^4$ K) such as HI, Mg II, Si II, Si III, and Si IV, probes of warm-hot gas (C IV, N V, and O VI, for $T \sim 10^5 - 3 \times 10^5$ K), as well as gas at the virial temperature of the halo (O VII; $T \sim 10^6$ K).  We will show that these comparisons make for a sensitive (and challenging!) probe of the phase of the inflowing and outflowing gas.  These runs are conducted on identical initial conditions, systematically varying the stellar feedback prescription to demonstrate their effect on the column density distribution of absorbers in the galactic halo.  In Section~\ref{method}, we describe the details of our simulations and our post-processing steps on both the numerical and observational datasets.  Section \ref{results} presents the results of these simulations and the consequent analyses.  We discuss these results and their implications for the nature of outflows, feedback and for use as a probe of different feedback models in Section \ref{discussion}.  Lastly, we summarize these conclusions in Section \ref{conclusions} and discuss future directions for subgrid models to match these added observational constraints.

\section{Methodology}
\label{method}

\subsection{Simulation Code}
\label{simulation}
The numerical methodology and initial conditions used in this paper are very similar to those employed in \citet[][hereafter Paper~I]{2012ApJ...749..140H}, and so we only briefly describe them here.  We perform all of our simulations using the adaptive mesh refinement (AMR) hydrodynamics code, Enzo\footnote{http://enzo-project.org} \citep{Bryan:1997wy, OShea:2004tr}. Enzo uses AMR to dynamically restructure the grid over the course of the simulation to achieve higher resolution in regions where predetermined refinement criteria are met.  It treats dark matter and stellar populations as collisionless particles modeled with an N-body particle-mesh gravity solver \citep{1985ApJS...57..241E}.  The equations of gas dynamics are solved using the ZEUS hydrodynamics engine \citep{Stone:1992iv}.

A minimum pressure floor is included in these simulations in order to keep the Jeans length resolved across all grid cells \citep{Truelove:1997bj}.  We follow the implementation of \citet{Machacek:2001fq}, wherein the ratio of the Jeans length to cell size is maintained at $J = L_{\rm J} / \Delta x \ge 8$ by the addition of an artificial pressure.

We model the process of star formation following a prescription similar to that in \citet{Cen:1992ju}.  A cell will spawn a star particle if the following criteria are met: (1) the cell's gas overdensity exceeds $\delta_{\rm SF}$, (2) the local velocity divergence is negative, and (3) the cooling time is less than the dynamical time for the gas cell.  When these conditions exist, a star particle with mass $m_{*}$ is created according to the relation:
\begin{equation}
m_{*}=\epsilon_{\rm SF}\frac{\Delta t}{t_{\rm dyn}}\rho_{\rm gas}\Delta x^3
\label{eq:sfr}
\end{equation}
where $\epsilon_{\rm SF}$ is the star formation efficiency per dynamical time;  $\Delta t$ is the local timestep; $t_{\rm dyn}$ is the local dynamical time, defined as $t_{\rm dyn} = (3 \pi / 32 G \rho)^{1/2}$; $\rho_{\rm gas}$ is the gas density; and $\Delta x$ is the grid cell width.  To prevent a large number of low mass particles, we adopt a minimum star particle mass of $m_{*,\rm min}$.  If $m_{*} < m_{*,\rm min}$, the star is formed with a probability of $m_{*}/m_{*,\rm min}$, and its resulting mass is the minimum of m$_{*,\rm min}$ and 90\% of the mass in the gas grid element \citep[e.g.][]{Tasker:2006kp}.  For these simulations, we set m$_{*,\rm min}$ = $10^5$ \msun.  We find that $\epsilon_{\rm SF} = 10^{-2}$ and $\delta_{\rm SF} = 10^3$ approximately reproduces the Kennicutt-Schmidt Law (see Paper~I).  This value of $\delta_{\rm SF}$ is equivalent to 0.1 cm$^{-3}$ at $z = 3$.

\subsubsection{Supernova Feedback}

Our implementation of stellar feedback includes a prescription for prompt injection of thermal energy from type II supernovae to the surrounding medium.  Individual star formation events produce star particles of $m_{*} \sim 10^{4-5}$ \msun, each representing an entire stellar population.  The high-mass stars within this stellar population will become supernovae and send energy, mass and metals into the local medium.  Because a stellar population is rarely created instantaneously in nature, we ``activate'' a small component of the star particle's mass each timestep to represent the stars that were effectively created during that period.  We parameterize the extended timescale over which this occurs with:
\begin{equation}
m_{\rm FB} = m_* ~\frac{\Delta t}{\tau}~ \frac{t-t_0}{\tau} ~ e^{-(t-t_0)/\tau}
\label{eq:feedback}
\end{equation}
where $m_{\rm FB}$ is the mass of newly activated stars for a given timestep, $t_0$ is the star particle creation time, and $\tau = \max(t_{\rm dyn}, {\rm 10\ Myr})$.  In order to prevent unrealistically short dynamical times, we apply a floor of 10 Myr to $\tau$.  We normalize our activated star mass so that the total mass released over the lifetime of the star particle is one-fourth the star's original mass (i.e. $\sum m_{\rm FB} = 0.25 m_{*}$).

The returned energy $e_{\rm FB}$ and returned metals $m_{\rm metals}$ due to stellar feedback are directly proportional to the mass of these new stars:
\begin{equation}
e_{\rm FB} =  m_{\rm FB} c^2 \epsilon_{\rm FB}
\end{equation}
\begin{equation}
m_{\rm metals} =  m_{\rm FB} ~ y
\end{equation}
where $c$ is the speed of light; $\epsilon_{\rm FB}$ is a feedback efficiency parameter; and $y$ is the fraction of metals returned relative to the activated stellar mass.  Thus at any timestep $\Delta t$ after the creation of a star particle, it returns $m_{\rm FB}$ mass, $e_{\rm FB}$ energy, and $m_{\rm metals}$ metals to the cell containing the star particle.  This thermal energy should heat the surrounding gas, increasing the Jeans mass, and prevent further immediate star formation.  

In the present investigation, we use a value of $y = 0.02$ consistent with previous work \citep{2011ApJ...731....6S}, and we include metal and mass feedback in all of our simulations.  In four of our simulations, we vary the thermal energy efficiency value: $\epsilon_{\rm FB} = 3 \times 10^{-6}$ (low), $1 \times 10^{-5}$ (medium), and $3 \times 10^{-5}$ (high).  These different energy efficiencies correspond to one $10^{51}$ erg SN for every 180, 60, and 18 \msun ~stars formed respectively, effectively sampling the range of feedback efficiencies used by other studies on this topic (see discussions in Section \ref{prev_work} and Paper~I).  

\subsubsection{Cooling Suppression} 
One of the shortcomings of the thermal feedback model is that thermal energy dumped into an underresolved volume will fail to bring that gas up to a high enough temperatures to prevent it from immediately radiating away this energy.  To address this, many researchers have adopted an {\it ad hoc} solution, to turn off radiative cooling in the gas for a period after a star formation event \citep{Gerritsen:1997tf, 2000ApJ...545..728T, SommerLarsen:2003eq, Stinson:2006id, Governato:2007dq}  This model is justified as an application of the Sedov-Taylor solution for a supernova blast wave applied at scales below the spatial resolution of the simulation, although these arguments are not entirely physically consistent (see Paper~I).  However, cooling suppression seems to effectively quench star formation and slow the overcooling problem particularly when combined with an additional feedback prescription \citep[][Paper~I]{Colin:2010ee, 2011MNRAS.410.1391A, 2011ApJ...742...76G, 2011MNRAS.410.2625P}.

Our implementation of cooling suppression is relatively simple.  After a star-formation event, we shut off cooling for a period $t_{\rm supp}$ in the gas cell containing our newly-formed star particle.  We allow this effect to track the star particle until $t_{\rm supp}$ has passed, just as our thermal prescription for feedback tracks the star particle for a time after its creation.  In this study, we set $t_{\rm supp} = $ 50 Myr, a value consistent with previous studies \citep[][Paper~I]{Stinson:2006id, Colin:2010ee}.

\begin{table*}
\begin{center}
    \begin{tabular}{|c|c|c|c|c|c|c|c|c|c|}
    \hline
    Name & Description & $M_{\rm 200}$ & $M_{\rm DM}$ & $M_{\rm star}$ & $M_{\rm gas}$ [hot, cold]\footnotemark[1] & $r_{200}$ & $v_{\rm circ,max}$ & $\epsilon_{\rm FB} $ & $ t_{\rm supp} $ \\
     & & ($10^{10}$ \msun) & ($10^{10}$ \msun) & ($10^{10}$ \msun) & ($10^{10}$ \msun) & (kpc) & (km s$^{-1}$) & & (Myr)\\
    \hline

    V & Vanilla & 110 & 93  & 17  & 2.5  [1.8, 0.6]  & 210 & 590 & 0 & 0 \\ [1ex]
    LF & Low Feedback & 110  & 92  & 15  & 3.6  [2.8, 0.8] & 210 & 560 & $3\times10^{-6}$ & 0 \\[1ex]
    MF & Medium Feedback & 110 & 91 & 11 & 5.1 [4.4, 0.8] & 210 & 520 & $1\times10^{-5}$ & 0 \\ [1ex]
    HF & High Feedback & 90  & 84  &  2 & 4.3  [3.0, 1.3] & 200 & 200 & $3\times10^{-5}$ & 0 \\[1ex]
    &Medium Feedback \& & & & & & & & & \\ [-2ex]
    \raisebox{1.5ex}{MFCS} & Cooling Suppression & \raisebox{1.5ex}{100}  & \raisebox{1.5ex}{90}  & \raisebox{1.5ex}{7}  & \raisebox{1.5ex}{6.6  [5.7, 0.9]}  & \raisebox{1.5ex}{200} & \raisebox{1.5ex}{390} & \raisebox{1.5ex}{$1\times10^{-5}$} & \raisebox{1.5ex}{50} \\
    \hline
    \end{tabular}
      \footnotetext[1]{Hot/cold division: T $= 2\times10^{4}$ K} 
\caption{We present each of our simulations, their simulation parameters, and the resulting halo characteristics at $z=0$.}
\label{tab:sims}
\end{center}
\end{table*}

\subsubsection{Radiative Heating and Cooling} 

In modeling the thermal state of the gas in these simulations, Enzo accounts for photoelectric heating from a metagalactic ultraviolet background and radiative cooling from metals present in the gas.  Because directly modeling the fractional components in each ionic state of every element is too computationally costly for simulations, we adopt a different tactic.  We employ the method pioneered by \citet{2008MNRAS.385.1443S} and subsequently used by other groups in a variety of contexts \citep{2009MNRAS.398...53B}, wherein all of the phases of H and He are followed individually throughout the simulation, but for more massive elements, a multi-dimensional lookup table is consulted to determine their aggregate contribution to heating and cooling using a single all-encompassing metallicity field.  All phases of H and He (H I, H II, He I, He II, He III, and e$^{-}$) are calculated by a non-equilibrium, primordial chemistry network, which uses a backwards-differencing formula to account for collisional ionization, recombination, brehmstrallung, Compton cooling, photoexcitation and photoionization rates in these species \citep{Abel:1997jz, Anninos:1997in}.  The metal heating/cooling lookup table is compiled using version 07.02.01 of the spectral synthesis code, Cloudy\footnote{http://www.nublado.org} \citep{1998PASP..110..761F}.  The table consists of five dimensions, one each for density, temperature, metallicity, electron fraction, and redshift.  The redshift index is included to determine the intensity and spectrum of the metagalactic ultraviolet background at that time, which we activate at $z = 7$ and henceforth follow the prescription set in \citet{Haardt:1996fq} as updated in this version of Cloudy.  Prior to reionization at $z = 7$, metal heating and cooling are only due to collisional ionization.  

\subsection{Initial Conditions} 
\label{initial_conditions}

In this study, we adopt the WMAP 5-year results \citep{Komatsu:2009ex}, specifically adopting $\Omega_{0} = 0.258$, $\Omega_{\Lambda} = 0.742$, $\Omega_{\rm baryon} = 0.044$, $\sigma_8 = 0.796$, $H_{\rm o}$ = 71.9 km s$^{-1}$.   Using the \emph{inits} program included in the Enzo distribution, we generate a $128^3$ particle grid over our $20h^{-1}$ Mpc volume and carry out a low-resolution simulation to identify a dark matter halo with mass $M_{200} \sim 10^{12}$ \msun~(see Paper~I for more details -- we use halo 26 of that paper).  Once the halo is identified, we select all particles within 2 virial radii at $z = 0$, and trace them back to their original positions at $z = 99$.  The region containing all of these particles at $z = 99$ is further refined with two nested refinement regions, each with 2-times better spatial resolution and 8-times lower mass dark matter particles, resulting in $M_{\rm DM} = 4.9 \times 10^6$ \msun~dark matter particles in the vicinity of our target halo.  Additional (adaptive) refinement occurs in a cell when its gas mass (or dark matter mass) exceeds  $4 \times 10^6$ \msun~(or $1.9 \times 10^7$ \msun~for the dark matter), resulting in a further increase of the spatial resolution by a factor of two, up to nine levels of refinement, achieving a maximum resolution of 425 co-moving pc.  The timescales over which our particles of different mass (i.e. stellar and dark matter) suffer from relaxation effects is on the order of a Hubble time, assuring these spurious effects are avoided.

\subsection{Description of the Simulations} 
We performed five different simulations using identical initial conditions and simulation parameters as described above.  The way in which these simulations differed was simply in their supernova feedback and cooling suppression parameters.  These runs include: a control run with no energetic feedback or cooling suppression [``Vanilla'']; three runs with no cooling suppression but varying feedback efficiencies of $\epsilon_{\rm FB} = 3\times10^{-6}, 1\times10^{-6},$ and $3\times10^{-5}$ [``Low Feedback'', ``Medium Feedback'', and ``High Feedback'', respectively]; and a run including cooling suppression and feedback with $\epsilon_{\rm FB} = 1\times10^{-6}$ [``Medium Feedback and Cooling Suppression''].  These runs and some of the properties of their halos at $z=0$ are listed in Table \ref{tab:sims}.  The simulations produced in this study have identical initial conditions and physical parameters to those associated with run H26SPM (i.e. the canonical run) in Paper~I save for the treatment of feedback and the presence of metals.  See that work for more detail.  It is worth mentioning that we do not tune our models in any way to produce the results herein, all parameters being either deduced from theoretical calculations or values in the literature.  Unless otherwise noted, all units herein are in comoving units, where $h^{-1}$ is excluded.

\subsection{Analysis Code} 
\label{analysis}
We applied a series of post-processing analyses to each simulation so as to render synthetic observations comparable to observational datasets for these simulated galaxies.  Each simulation outputs a datafile to disk every 10 Myr, representing the state of the gas, stars and dark matter at that time.  Following the method described in \citep{2011ApJ...731....6S}, we processed each datafile to compute the spatial distribution of various different ionic species.  The density of the $i$th ionic species of a given element, $X$, is given by 
\begin{equation}
n_{X_i} = n_{\rm H} \times  (n_{X}/n_{\rm H}) \times (n_{X_i}/n_{X})
\end{equation}
where $n_{\rm H}$ is the density of hydrogen, $(n_{X}/n_{\rm H})$ is the elemental abundance of $X$ and $(n_{X_i}/n_{X})$ is the ion fraction in the $i$th energized state.  We internally track the neutral and excited states of hydrogen and helium throughout the simulation, but we represent all metals as a single metallicity field.  Thus, we assume the metallicity of any given metal $X$ tracks with this overall metallicity $Z$ consistent with solar abundance ratios: $n_{\rm X} / n_{\rm H}$ = $Z \times (n_{\rm X} / n_{\rm H})_{\odot}$.  To calculate $(n_{X_i}/n_{X})$, we create a lookup table using the spectral synthesis code, Cloudy (version 07.02.01), which gives $(n_{X_i}/n_{X})$ uniquely determined by the gas temperature, gas density, and the photoionizing spectrum and intensity at that time.  We use the redshift-dependent ultraviolet background included in Cloudy as the photoionizing source \citep[a 2005 updated version of][]{Haardt:1996fq}, which takes into account reionization by QSOs and galaxies.  The Cloudy lookup table accounts for collisional ionization as well as photoionization by the metagalactic UV field (although we do not track local photoionizing sources).  This infrastructure coupled with the yt simulation analysis suite\footnote{http://yt-project.org} \citep{Turk:2011dd}, allows us to add any arbitrary ion and element to our simulation as an additional data field, as defined by the local baryon density, metallicity, temperature, and redshift.  We process all of our data outputs to include fields for the astrophysically important ions of H I, Mg II, Si II, Si III, Si IV, C IV, N V, O VI, and O VII.

For each data output of each simulation, we locate the galaxy of interest, and create a series of ten column-density projections centered on the galaxy from randomly oriented vectors passing through the galaxy in each of the selected fields.  Each projection consists of two 128$^2$-pixel images, one with a field of view of 2 Mpc and the other with 100 kpc, allowing us to achieve a large dynamical range in resolution (1 Mpc to 1 kpc resolution in the central region of the galaxy).  These images provide a map of the column density of each species and can be compared directly against observations.  Figure \ref{fig:projections} displays a few such projections from the MFCS run in H I, C IV, O VI and total gas density at $z = 0.5$.  For illustration's sake, these images are 500 kpc on a side, with a circle overplotted representing $r_{200}$, the radius at which the galactic density drops to 200 times the mean density of the universe.  The column density limits chosen are those used later in the radial profiles for each species.

For the image plane, pixels are identified by their comoving distance from the central pixel.  These pixel radii are then binned, and the bins are populated by the column density maps of a series of images for a given galaxy.  The median column density value in each radius bin is computed to produce a radial profile for that galaxy, representing the typical column density one might expect to detect at a given radius were one to observe it from a random angle.  We found early on that there were significant temporal peculiarities (e.g. infalling satellites) which made it difficult to compare one run at a given redshift against another, so we made radial profiles consisting of data outputs over a range of redshifts to wash out this effect.  To give some insight into the range of column densities at each radius, we also include some quartile column density ranges (i.e. 25\% and 75\%) in these radial profiles.  

For the physical quantities of metallicity and temperature, we take a mass-weighted average projection for each quantity as opposed to a column density.  All further processing steps on these projections follow the same prescription as for the atomic species projections.

\subsection{Processing the Observational Datasets} 
\label{observations}
In order to say something meaningful about our numerical models, we compare them directly against observations taken in a similar manner.  Typically, gas in these species is not observed in emission, but rather in absorption due to its low density (emission $\propto n^2$, whereas absorption $\propto n$).  We identified observational studies which correlate the proximity (i.e. $\rho \equiv$ impact parameter) of absorbers to nearby host galaxies for a subsample of our chosen atomic species.  The observational studies against which we compare our data are:  H I (Ly$\alpha$) from \citet{2011ApJ...740...91P} \citep[utilizing data from][]{2004ApJS..152...29P, 2008ApJ...679..194D, 2008ApJS..177...39T}; Mg II: \citet{Chen:2010ir}; C IV from \citet{Chen:2001hj}; O VI from \citet{Tumlinson:2011wm}; and O VII from \citet{2008ApJ...672L..21Y}.  

In two of these studies \citep[e.g.][]{Chen:2001hj, Chen:2010ir}, the absorption features were not fully resolved into individual components, and so only equivalent widths of the absorbing gas were provided, whereas in all of the other studies both equivalent width and column densities of the absorber are reported.  In order to directly compare our results against all of these studies, we needed to convert any equivalent widths to column densities.   Converting between equivalent width and column density is not a trivial proposition, as it requires knowledge of the temperature and velocity distribution of the gas containing the absorbing species.  To make an approximate translation, we used a ``curve of growth'' analysis.  We applied a fitting function relating equivalent width and column density as found in \citet{Draine:2011tr}:
\begin{equation} \label{eq:fit}
W \approx \left\{
\begin{array}{l r}
\sqrt{\pi} \frac{b}{c} \frac{\tau_0}{1 + \tau_0 / (2 \sqrt{2})} & \mbox{$\tau_0 <$ 1.254} \\
\left[\left(\frac{2b}{c}\right)^2 \ln{\left(\frac{\tau_0}{\ln{2}}\right)} + \frac{b}{c} \frac{\gamma \lambda}{c} \frac{(\tau_0 - 1.254)}{\sqrt{\pi}}\right]^{1/2} & \mbox{$\tau_0 >$ 1.254}
\end{array}
\right.
\end{equation}

\begin{equation}
\tau_0 \approx \sqrt{\pi} \frac{e^2}{m_e c} \frac{N_l f_{lu} \lambda_{lu}}{b}
\end{equation}
where $W$ is the equivalent width of a given absorption line for a transition from state $l$ to state $u$, $b \equiv \sqrt{2}\sigma_{V}$ is the velocity dispersion of the absorbing gas, $c$ is the speed of light; $\tau_0$ is the optical depth at absorption line center, $\gamma$ is the intrinsic width of the absorption line, $\lambda$ is the wavelength of the transition, $N_l$ is the column density of material in the $l$ state, and $f_{lu}$ is the oscillator strength of the transition.

This relation converts equivalent width to column density, but we want the inverse, so we numerically inverted the function applying the bounds-bisection method \citep{num_recipes}.  We used \citet{1996ADNDT..64....1V} for accurate oscillator strength, wavelength, and $\gamma$ values for the various transitions.  

To deduce the appropriate value of $b$ for our absorber populations, we utilized the equivalent width and column density data in \citet{Tumlinson:2011wm} and \citet{2011ApJ...740...91P}.  Applying equation \ref{eq:fit} to calculate column density from these equivalent widths for a variety of $b$-values, we then made a $\chi$-squared fit with the listed column density values to ascertain the best choice of $b$ for O VI and H I respectively.  Our results indicated $b_{\rm O VI} =~$58 km~s$^{-1}$ and $b_{\rm H I} =~$29 km~s$^{-1}$.  Since the gas responsible for absorbing C IV is in a similar temperature regime (slightly lower) to that of O VI absorbing gas, we used $b_{\rm C IV} \approx b_{\rm O VI} = 58$ km~s$^{-1}$.  Additionally, Mg II has a similar ionization energy to H I, and Mg II is thought to trace H I at $N_{\rm HI} > 10^{16}$ cm$^{-2}$ \citep{Putman:2012vp}, so we let $b_{\rm Mg II} \approx b_{\rm H I} = 29$ km~s$^{-1}$.  Because the lines of these species may have multiple components, these $b$ values should not be take too literally, but for the purposes of this current comparison, they suffice.  

The uncertainties endemic to this type of conversion are mostly contained in our choice of $b$ value.  Changes in $b$ of 3\% result in changes in calculated column density of $\sim 3\%$ and $\sim25\%$ depending on whether the absorber is in the linear or saturated region of the curve of growth respectively.  Fortunately, the majority of our C IV (95\%) and Mg II (70\%) absorbers lie in the linear regime, with only the strongest absorbers populating the saturated region.  It is worth noting that we are only applying this conversion to two observational datasets (Mg II and C IV) for comparison's sake, and not to our own numerical datasets, so any problems with this conversion have limited repercussions to the study at hand.  Furthermore, the main results of this paper come from comparisons against the O VI dataset, where this conversion was not required.  We thus convert the observational datasets to approximate column densities (and proper units to comoving units) for comparison against the numerical results.

\section{Results}
\label{results}
Our suite of simulations yields halos with similar bulk properties, consistent with the fact that they all share the same initial conditions.  We catalog the mass of each component (dark matter, gas, and stars) and the maximum circular velocity for the primary halo in each of our simulations at $z = 0$ in Table \ref{tab:sims}.  The mass in each component is measured interior to $r_{200}$ at $z = 0$.  There is little variation in the dark matter and total mass across our simulations except in the case of the high feedback run where large baryonic outflows have softened the potential and diminished the overall mass of the system.
    
\begin{table}
\begin{center}
\footnotesize{
    \begin{tabular}{|c|l|c|c|}
    \hline
    Species & $E_{\rm ion}$ & Redshift & Reference\\
    \hline
    H I & 13.6 eV & 0.05 - 0.2 & \citet{2011ApJ...740...91P}\\
    Mg II & 15.0 & 0.1 - 0.5 & \citet{Chen:2010ir}\\
    Si II & 16.3 & --- & ---\\
    Si III & 33.5 & --- & ---\\
    Si IV & 45.1 & --- & ---\\
    C IV &  64.5 & 0.08 - 0.9 & \citet{Chen:2001hj}\\
    N V & 97.9 & --- & ---\\
    O VI & 138. & 0.1 - 0.35 & \citet{Tumlinson:2011wm}\\
    O VII &  739. & 0 & \citet{2008ApJ...672L..21Y}\\
    \hline
    \end{tabular}
    }
\caption{The observational datasets used to compare against our numerical radial profiles. We also include ionization energies for our selected ionic species. }
\label{tab:observations}
\end{center}
\end{table}

\begin{figure*}
\begin{center}
\includegraphics[scale=0.8]{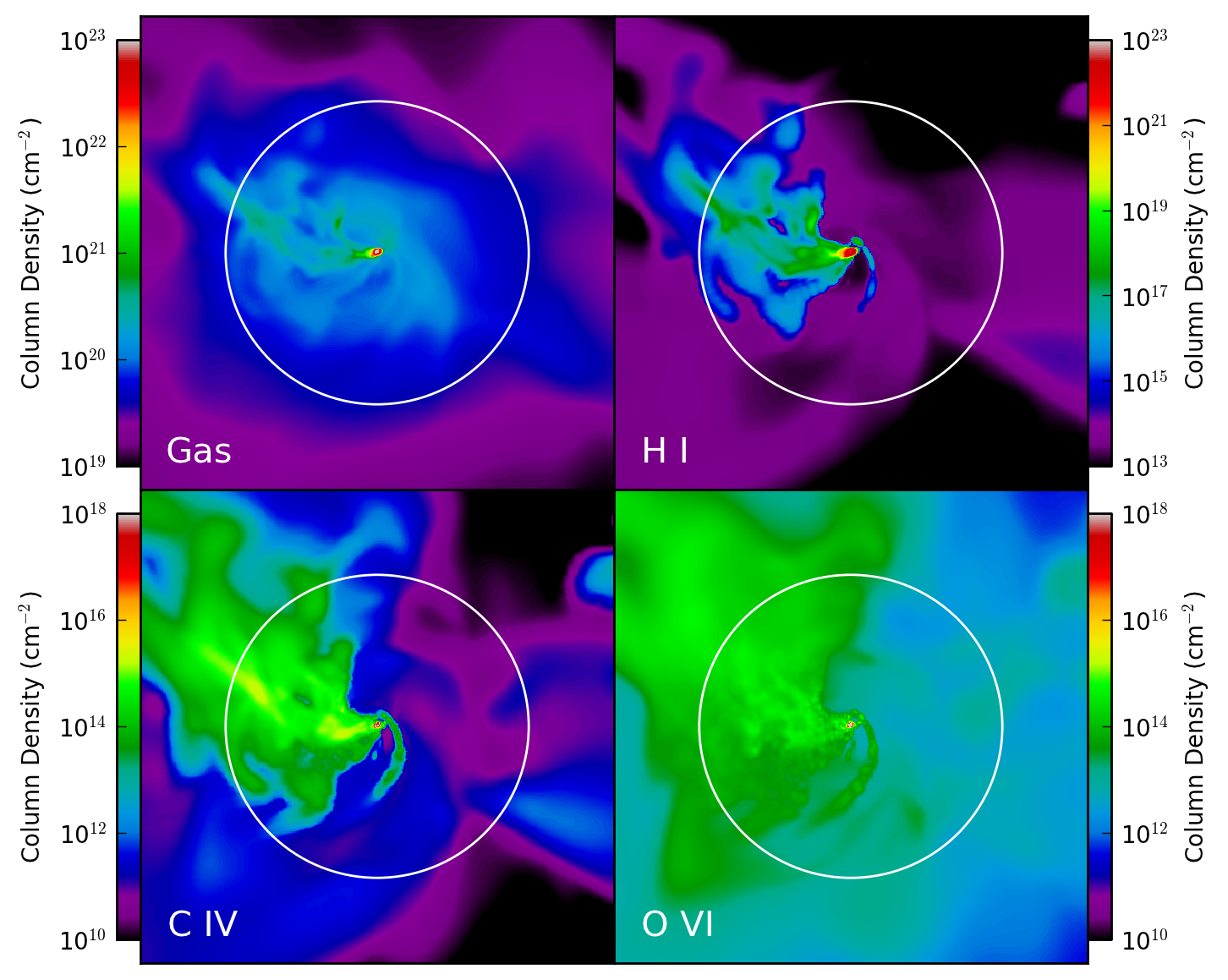}
\end{center}
\caption{Sample projections from the medium feedback and cooling suppression run at $z=0.5$.  Each image represent a column-density map of a region with 500 comoving kpc on a side.  The white circle denotes $r_{200}$ = 160 kpc.}
\label{fig:projections}
\end{figure*}

In Paper~I, we examined the rotation curves and star formation histories of a set of simulations very similar to those used in this paper.  Here, we focus on the observable properties of the circumgalactic medium (CGM) at low redshift -- in particular, we report on the column density distributions of a wide variety of atomic species, observable in absorption against background quasars.  We initially restrict ourselves to low redshift, as there exists a substantial body of observations which link absorption strength to projected distance from the galaxy.  In the following sections, we first examine the column density distributions, then explore the physical properties of the gas giving rise to this absorption, and finally investigate the redshift evolution of the predicted absorption strength.

\subsection{Radial Profiles for Atomic Species}
In this study, we examine the radial column density distribution for a number of absorption-line-generating species, well-sampling the range of ionization energies (and thus temperature/density regimes) of the CGM as shown in Table \ref{tab:observations}.  At the two ends of the spectrum, the H I traces the cold, dense clouds, whereas the O VII probes the hot coronal gas.  Between these regimes is a continuum of layers of material acting as the interface between these two phases.  This layering effect takes place on many different spatial scales from small star-forming clumps of cold gas up to the galactic scale.  Figure \ref{fig:projections} demonstrates this at the latter scale, where the cold H I of the inner halo is enshrouded in warmer C IV-bearing gas, which is further encased and extended by gas exhibiting O VI.  This provides us with some intuition for understanding the following galactic profiles for these species.

\begin{figure*}
\begin{center}
\includegraphics[scale=0.78]{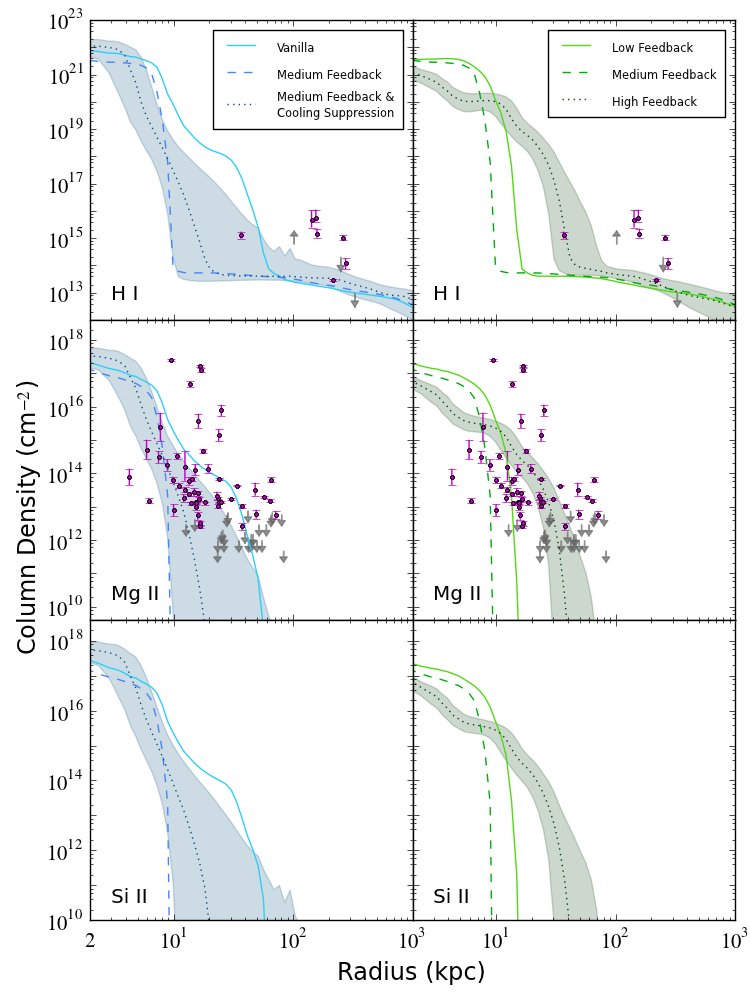}
\end{center}
\caption{Radial profiles for column densities of H I, Mg II, and Si II in comoving units covering the period $z = 0.1 - 0.5$.  The color bands represent the quartile values ($\pm25\%$) overplotted on the median for two of the models: MFCS and HF.  Also shown are observational datasets with detections as magenta points and limits in gray arrows from the datasets listed in Table \ref{tab:observations}. The numerical datasets contained in Figures \ref{fig:radprof2}, \ref{fig:radprof3}, \ref{fig:radprof4}, and \ref{fig:radprof1} are available in tabular form for public download at http://chummels.org/CGM.html .}
\label{fig:radprof2}
\end{figure*}

\begin{figure*}
\begin{center}
\includegraphics[scale=0.78]{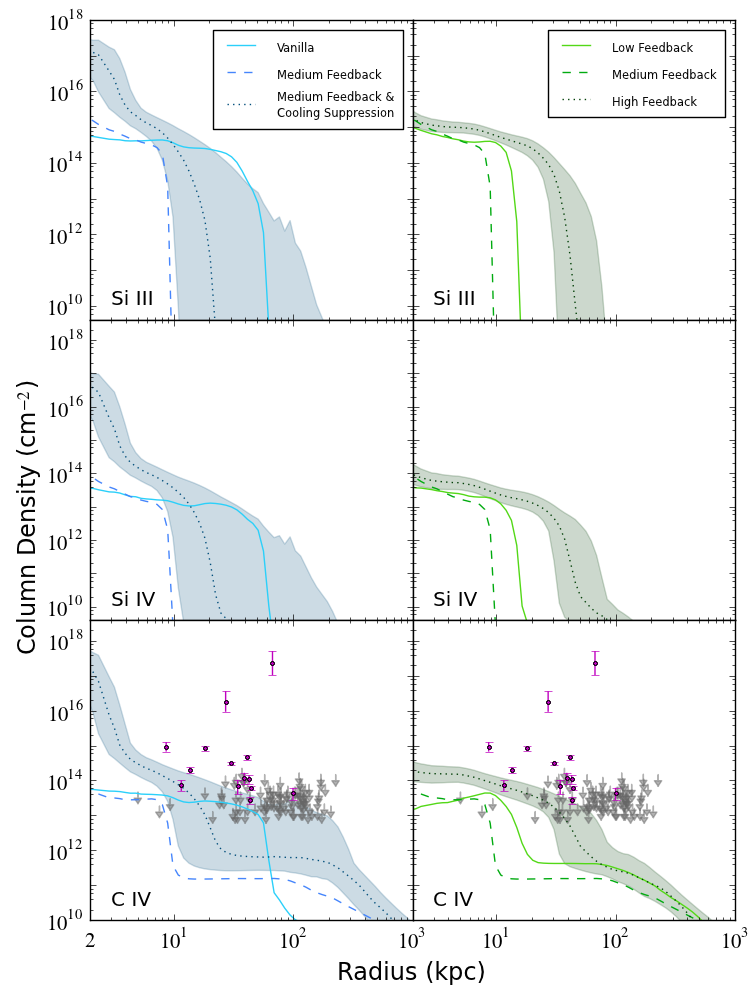}
\end{center}
\caption{Similar to Figure \ref{fig:radprof2} but for Si III, Si IV, and C IV.}
\label{fig:radprof3}
\end{figure*}

\begin{figure*}
\begin{center}
\includegraphics[scale=0.78]{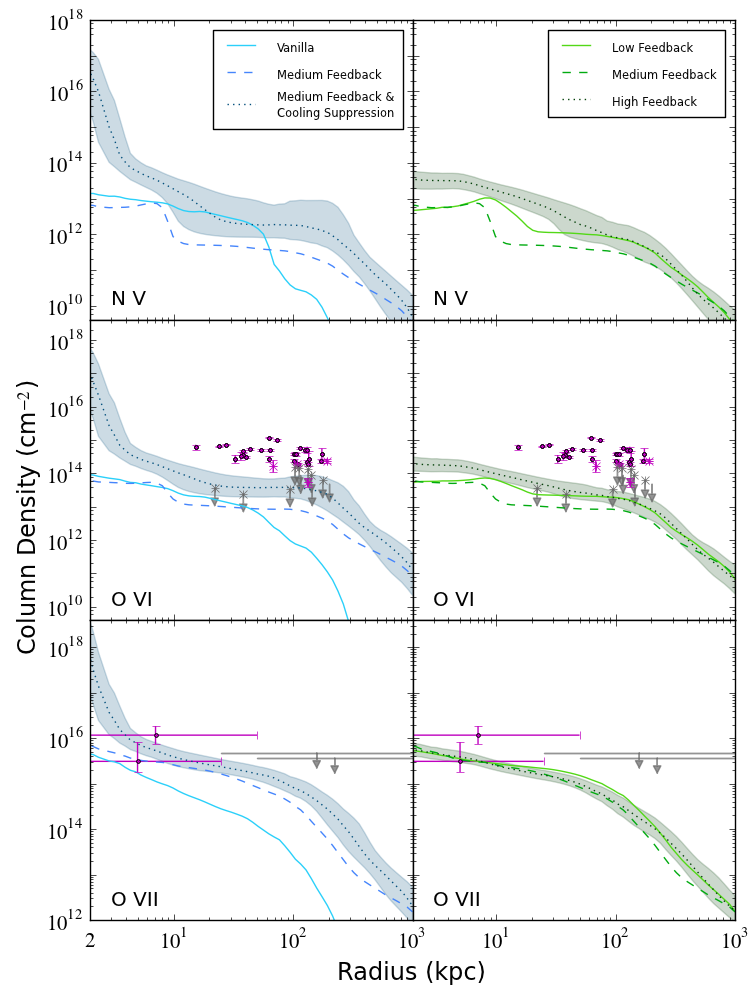}
\end{center}
\caption{Similar to Figure \ref{fig:radprof2} for N V, O VI, and O VII.  The O VI dataset \citep{Tumlinson:2011wm} is split into star-forming galaxies (circle symbols and plain arrows) and passive galaxies (``asterisk'' symbols and ``asterisk'' arrows).}
\label{fig:radprof4}
\end{figure*}

We produce radial profiles of the median gas column density in nine different ionic species as a function of projected radius extending from 2 kpc to 1 Mpc for our simulated galaxies.  We probe this gas from the disk-dominated region ($r < r_{\rm inner} \sim 20$ kpc) through the CGM ($r_{\rm inner} < r < r_{200}$) into the intergalactic medium ($r > r_{200}$).  Figures \ref{fig:radprof2}, \ref{fig:radprof3}, and \ref{fig:radprof4} present these results (Figure \ref{fig:radprof1} shows total gas column densities for comparison).  The left and right sides of each figure show the radial profiles of the same quantity but for different simulations.  The left sides display a comparison of simulations with and without different forms of feedback: V, MF and MFCS, whereas the right sides exhibit a progression of thermal feedback intensities: LF, MF and HF (the MF simulation is repeated on both sides to ease comparison).  Each profile gives the median value expected in an observation of a galaxy at a given radius over the redshift range $z = 0.1 - 0.5$ (selected to approximately match the redshift ranges of the observations).  To give some idea of the variation around the median absorber strength, in each panel we overplot a color band representing the quartiles of the column density distribution (i.e. 25\% to 75\%) for one model: MFCS (left) and HF (right).  The quartile distribution associated with the HF model is most indicative of the spread of all the other models, whereas MFCS produces a particularly spread-out distribution suggesting more multiphase substructure.  
In Figure \ref{fig:radprof2} we present radial profiles for our lowest ionization energy species, H I, Mg II, and Si II; Figure \ref{fig:radprof3} shows medium energy ions, Si III, Si IV, and C IV; and Figure \ref{fig:radprof4} displays the highest energy species N V, O VI, and O VII.  Overplotted on several profiles are low-redshift observational datasets, where magenta points represent detections and grey arrows mark limits.  These datasets, their sources, and their redshift ranges are listed in Table \ref{tab:observations}.

The H I, Mg II, and Si II datasets (the H I group) in Figure \ref{fig:radprof2} demonstrate similar behavior across all of our models, which is consistent with the fact that their ionization energies are very close to each other (see Table \ref{tab:observations}).  All of the simulations show a high-column-density core (i.e. the disk) in these species out to $r\sim10$ kpc, at which point the profiles drop to trace levels by 10-50 kpc (i.e. the extent of cool gas surrounding the disk).  This transition radius, $r_{\rm inner}$, is related to the extent of the galactic wind in that it marks the outer boundary of this drop, where the material falls off to the $\sim 10^{13}$ cm$^{-2}$ level.  Each feedback model predicts a different $r_{\rm inner}$, sometimes differing over 40 kpc in radial location, identifying $r_{\rm inner}$ as a potentially useful differentiator between feedback models.  Our V and HF simulations can match the bulk of the Mg II observations; however none of the models reproduce the most extended high column density observations in H I and Mg II.  Note that we have subsampled the H I observations \citep{2011ApJ...740...91P} including only those detections associated with galaxies in the $0.5\rm L^{*} < L < 2\rm L^{*}$ range since these best match the characteristics of our simulated galaxy.

In Figure \ref{fig:radprof3}, the medium-ionization species Si III, Si IV, and C IV (the ``C IV group") display somewhat similar behavior to those of the H I group, each exhibiting a core and precipitous drop in material around 10-40 kpc.  The floor level of the C IV absorption strength (beyond $r_{\rm inner}$) is larger than either of the other two lines, presumably due to the fact that it is tracing warmer gas.  We also compare to the observed C IV dataset, but note that this comparison is difficult due to the somewhat contradictory limits and detections in the outer halo; however, all of the simulations appear to be an order magnitude too low in column density in the probed radial regime ($r\sim10-200$ kpc).  

The tracers of warm and hot gas, N V, O VI, and O VII, have more extended and flatter distributions than the H I and C IV groups. The O VII distribution, which probes gas with temperature close to the virial temperature of the halo is particularly flat (note the different vertical scale).  Our models that best fit these observational datasets are the HF and MFCS simulations; however even those profiles are an order of magnitude below the observations of O VI in the $r\sim10-100$ kpc range.  These same models reasonably fit the O VII observations, although this dataset has relatively weak constraints.  The vanilla run conspicuously falls off in column density at $r\sim50$ kpc in all of the species (we will address the reason for this in more detail below).  The O VI observations are separated into ``passive'' galaxies (plotted with an asterisk) and star-forming galaxies (plotted without an asterisk) \citep[see][for a full description]{Tumlinson:2011wm}.  While one would expect the active galaxies from the observational sample to be more like our simulated star-forming disk galaxy, it appears that the passive systems yield a much better match.

\subsection{Radial Profiles for Physical Quantities}
\begin{figure*}
\begin{center}
\includegraphics[scale=0.78]{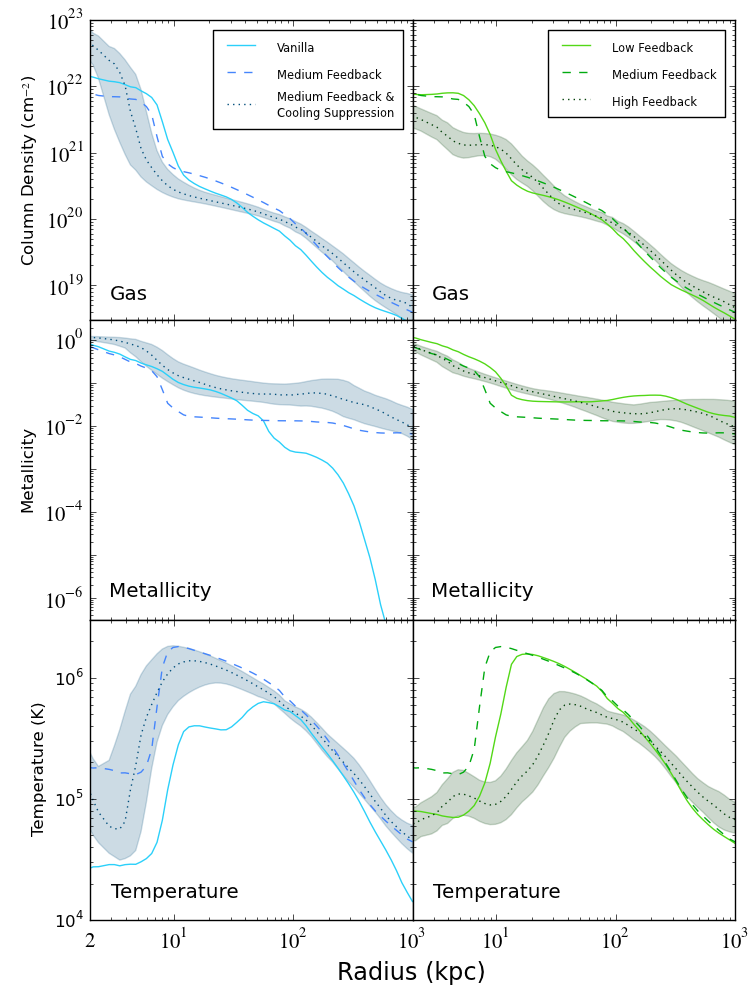}
\end{center}
\caption{Radial profiles for physical quantities: column density of gas, mass-weighted metallicity, and mass-weighted temperature covering the period $z = 0.1 - 0.5$. The color bands represent the quartile values ($\pm 25\%$) overplot on the median for two of the models: MFCS and HF.}
\label{fig:radprof1}
\end{figure*}

In order to better understand the physical basis for the simulation predictions, we next turn to the physical properties of the CGM gas itself.  We present radial profiles for median gas column density, median metallicity, and median temperature in Figure \ref{fig:radprof1} in a manner similar to Figures \ref{fig:radprof2}, \ref{fig:radprof3}, and \ref{fig:radprof4}.  In order to make a fairer comparison to the projected column density profiles, we compute these profiles using the same machinery, first carrying out several projections for each quantitiy (mass-weighting for the metallicity and temperature), and then computing the median value for each projected radius over the same $z=0.1$ to 0.5 redshift range used earlier.  The profiles computed in this way are similar to the three-dimensional quantities, but are more straightforward to compare at a given projected radius.

For the gas column density profile, there is not a great deal of variation (within a factor of 2) from one simulation to another, suggesting that different feedback models play only a modest role in rearranging the radial distributions of the gas mass within a galaxy.  

The metallicity profiles demonstrate a much more dramatic difference between those runs incorporating energetic feedback (LF, MF, MFCS, HF) and the run which does not (V).  While all of the models share roughly the same median metallicity in their inner 50 kpc (within 1 dex of Solar), the median metallicity of the vanilla run plunges to nearly primordial values beyond that distance, explaining why the vanilla run drops off in C IV, N V, O VI and O VII.  All of our simulations share the ability to produce and return metals to the surrounding medium from star-formation events, but only the ``feedback'' runs are capable of depositing energy along with those metals.  It is clear that the inclusion of some form of energetic feedback is necessary to mix and disperse high-metallicity gas beyond 50 kpc.  Lacking energetic feedback, the V simulation effectively has no outflow of material, and the metals get mixed up solely by gas motions in the inner halo region.  This fact makes it very useful for designating as an \emph{inflow-only} model, thus cleanly isolating the characteristics of infalling material versus models generating outflows.  Because our inflow-only simulation shows similar concentrations of gas in the outer halo, we can conclude that most of the gas in the halo must have been deposited there by inflows.  Evidently, the main effect of feedback is to mix in higher-metallicity material to large halo distances, but not to enhance the density there dramatically.

All of the simulations share the same general behavior in the projected temperature profiles: low in the core (the cold gas of the disk) out to $\sim r_{\rm inner}$, higher values extending to around the virial radius, and then decreasing beyond that.  Note that the ``cold" gas temperatures ($T \sim 10^5$ K)  seen in the core in these profiles reflect the presence of multi-phase gas.  The LF, MF and MFCS runs exhibit a high peak temperature of $T\sim20$ million Kelvin at $r_{\rm inner}$, where the high density of the galactic disk drops off.  Conversely, the V and HG runs peak farther out ($r\sim40$ kpc) at a lower temperature ($T\sim7\times10^5$ K).  The lack of clear trends in the temperature profiles indicate that low values in the center reflect simply how much cold gas survives in each model, while the temperature in the outer parts of the halo probably depends largely on the halo's virial temperature.  These temperature profiles, however, are clearly related to the species profiles shown earlier -- the low ionization species (particularly the H I group) only appear where the mass-weighted gas temperature is low.  Even the C IV group are correlated strongly with the extent of cold gas, probably indicating that it is formed in multi-phase gas.  Finally, the column density profiles of the higher ionization species are generally flatter than the gas profile, indicating that they are primarily coming from the outer halo and their slight overdensity in the core is mostly a projection effect.  This is particularly true of the O VII profile, which is very flat.

\subsection{Redshift Evolution of Radial Profiles}
\begin{figure*}
\begin{center}
\includegraphics[scale=0.78]{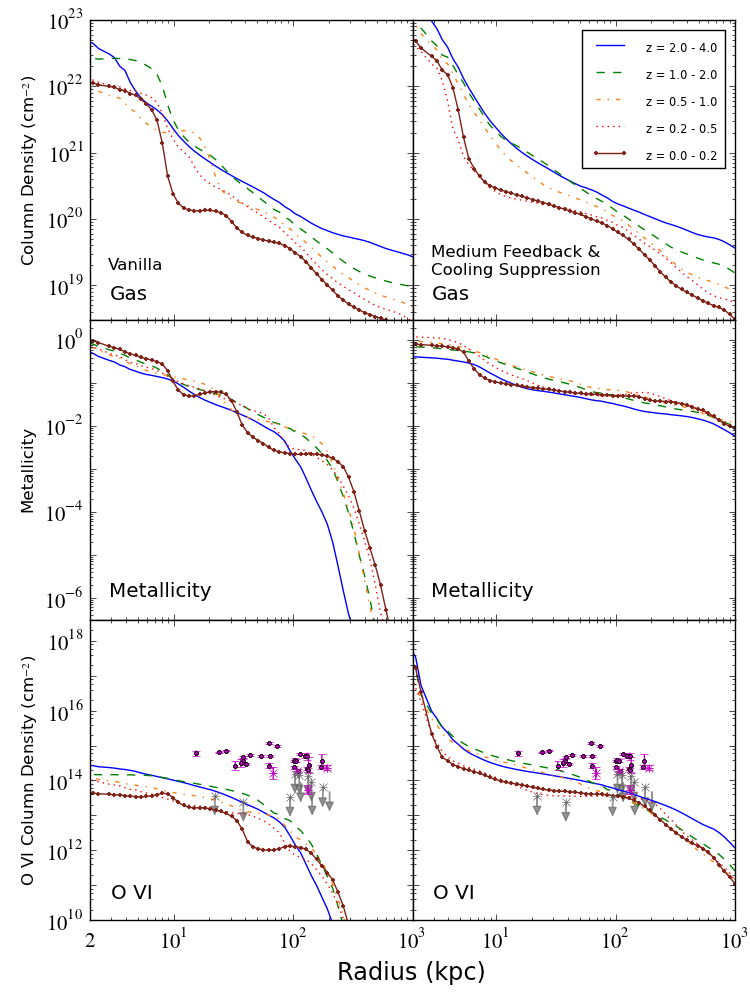}
\end{center}
\caption{Radial profiles for gas, metallicity and O VI as they change over time.  The left and right sides display the evolution of radial profile for the V and MFCS runs respectively.}
\label{fig:redshift}
\end{figure*}

There exists a significant body of observations as well as a growing number of computational studies investigating the radial profiles of metal-bearing gas in the CGM of galaxies at high redshift \citep[e.g.][]{2011ApJ...737L..37A, 2011MNRAS.411..826T, 2011MNRAS.418.1796F, 2011MNRAS.416.1723B, Shen:2012ux}.  While the focus of the current work is low redshift, it is worth examining how the CGM evolves over time in our simulations.  Figure \ref{fig:redshift} presents radial profiles of gas column density, mass-weighted metallicity, and O VI column density for two of our models (V and MFCS) as a function of redshift.  These profiles are constructed using the same method of median-sampling many data outputs over a range of redshifts so as to wipe out any temporal peculiarities in the profile.  The figure demonstrates trends in the behavior of certain quantities with time sampling redshifts from $z$ = 4 to $z$ = 0.  

\begin{figure*}
\begin{center}
\includegraphics[scale=0.6]{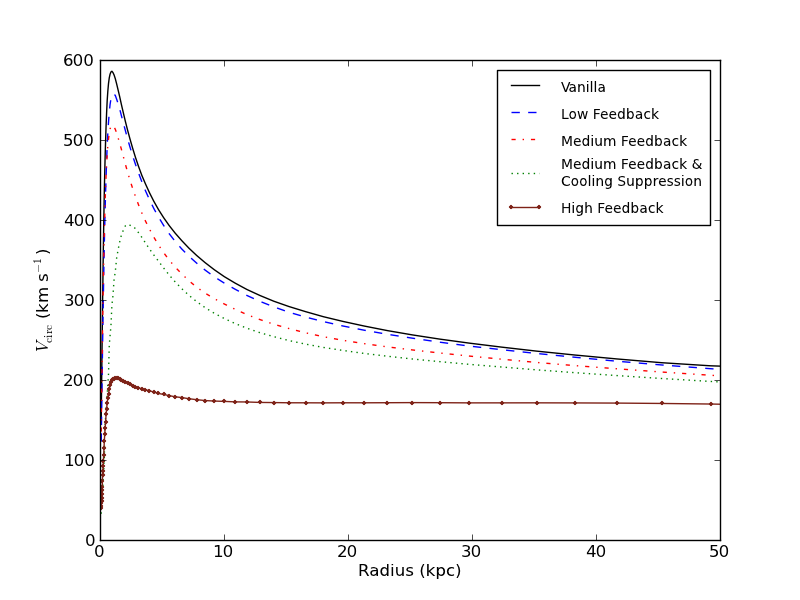}
\end{center}
\caption{Rotation curves for each of the simulations at $z = 0$.}
\label{fig:rotcurve}
\end{figure*}

The gas column density behaves similarly in both of our models, consistent with an expanding universe.  From an average redshift of $z=3$ to $z=0$, the scale factor changes by a factor of four, so that the column density of a uniform medium integrated over a constant comoving path length should decrease by a factor $N^2 = 16$, which is approximately consistent with these profiles, particularly at large projected radii.  On the other hand, metallicity appears to increase over time in both models, especially for the inflow-only simulation, although the effect is quite weak.  In the MFCS run, the metallicity grows with time at all radii, but by only a few percent.  Much of the star formation occurs at early times and is spatially distributed throughout the galaxy, quickly enriching the gas at all radii, so that subsequent star formation seems to do little to further increase the metallicity throughout the halo.

Finally, the O VI radial profile in this figure exhibits behavior similar to the gas profile, in that the column densities at all radii drop over time.  While the high-redshift version of the MFCS model matches the O VI observational data relatively well, once it has reached low redshift (the redshift of the actual observational dataset), it has fallen almost an order of magnitude.  This behavior of ionic column densities to decline by an order of magnitude at all radii is somewhat generic across all of our species, although some of the line/simulation combinations have a more subtle time-dependence belying the complex nature of the Saha equation.

\subsection{Rotation Curves}
In Paper~I, we analyzed the efficacy of different subgrid models (including feedback prescriptions) using the rotation curves of the resulting galaxies as a metric of success.  We revisit this analysis with the current models in Figure \ref{fig:rotcurve}.  Consistent with Paper~I, simulations V (canonical run), LF and MF produce galaxies with very peaked rotation curves ($v_{\rm circ}\sim600$ km s$^{-1}$), clearly suffering from the angular momentum problem.  The simulation including cooling suppression (MFCS) was successful at reducing this peaked value significantly to $v_{\rm circ}\sim400$ km s$^{-1}$.  The high feedback run resulted in a system with a very flat rotation curve and about 20\% less total mass than the other halos.  While our MFCS and HF simulations may appear as successful models (i.e. realistic) based on this metric alone, the preceding analysis comparing ionic column densities makes it clearer that they do not predict significantly different absorber distributions from the lower feedback runs (and fail to match observations).  This fact alone reveals that QSO absorption line observations are an important constraint for feedback models to match in addition to traditional measures of success (e.g. rotation curves, star formation rates, Tully-Fisher Relation, Kennicutt-Schmidt Relation).

\begin{figure*}
\begin{center}
\includegraphics[scale=0.8]{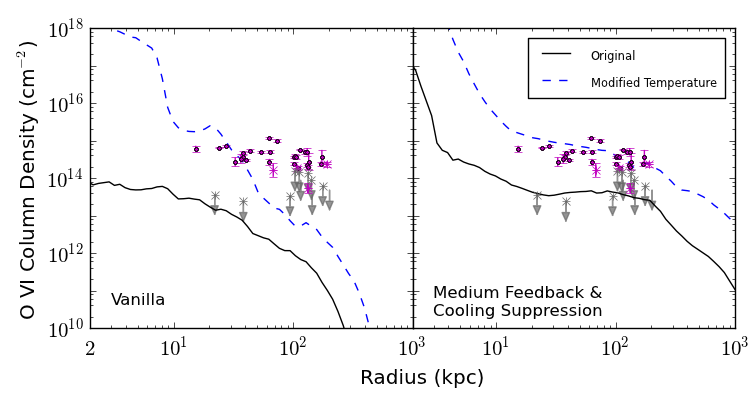}
\end{center}
\caption{Radial profiles of O VI column density for a single data output at $z=0.2$.  The left and right sides show the V and MFCS runs respectively. The O VI column density of the original model is displayed in black, whereas the maximum O VI column density when we leave temperature as a free parameter is displayed in blue.}
\label{fig:deduce}
\end{figure*}

\section{Discussion}
\label{discussion}
For all of the simulations, we see the same general behavior over each of the ionic radial profiles.  The low-energy species of H I, Mg II, Si II, Si III, Si IV, and C IV display a core column density level out to a characteristic radius, $r_{\rm inner}$, beyond which they fall steeply to a floor of trace levels indicating that these ions tracing cool gas are mostly associated with the disk.  As we increase the ionization energy of the species, the core column density diminishes and the floor column density level increases indicating warmer, more distributed gas being traced.  This trend continues to the highest energy ions of O VI and O VII, which display hot, extended gas in mostly flat radial profiles.

It is clear from our results that these simulations do not produce column densities distributions of various ionic species matching QSO absorption observations.  In general, our models predict column densities which are consistently at lower values and less extended than the observational studies, save for the predictions of the hottest gas in O VII absorption.  Furthermore, there is only a hint of a trend in our data suggesting a path to resolve this discrepancy (e.g. by increasing feedback efficiency).  So why do our models fail to agree with observations?

One obvious problem with our method is that we only consider column densities of material within 1 Mpc of our galaxy.  In reality, quasar sightlines pass through a much larger volume, and it is possible that additional intervening absorbers with velocities similar to our galaxy exist, which would increase the column densities in subsequent radial projections.  However, this effect is likely to be small, so we seek additional explanations for these column density discrepancies.

\subsection{Matching O VI}
In order to better understand the shortcomings of our models, we attempt to reconstruct the conditions necessary to reproduce the observational datasets.  In particular, we try to match the O VI observations, since of all the observational datasets compared, they have the smallest errors and are the least likely to be susceptible to inaccuracies from photoionization (see section \ref{limitations} below).  As noted previously, our radial profiles of O VI column density are 1-2 orders of magnitude lower than the observations in the $r\sim10-100$ kpc region.

Broadly speaking, one could imagine two explanations for this deficit.  First, there could simply be an insufficient amount of metal-enriched gas at the projected radii in question, or second, the gas could be there in sufficient quantities, but not in the correct ionization state, perhaps because the temperature is too high or too low.  O VI is particularly susceptible to the thermal state of the gas, as it only exists in sufficient quantities in a relatively narrow temperature range (assuming collisional ionization).  To explore which of these two explanations is most consistent with our simulations, we take the radial profiles of physical quantities generated in Figure \ref{fig:radprof1}, and systematically vary either the density, temperature, or the metallicity at each projected radius in order to maximize the resulting O VI column densities, holding all other quantities constant.   Under the assumption that collisional ionization of O VI dominates, we fix the radiation intensity (see Section \ref{limitations}).  

We test this out on a single data output at $z = 0.2$ in the MFCS run although the result holds at other times.  First, we find that for a fixed metallicity and temperature, changing the density has little effect -- this might be somewhat counterintuitive, but is expected because we are keeping the column density constant, and O VI collisional ionization depends very little on density.  The metallicity has a much larger effect, but we find that there is no solar or sub-solar metallicity values which will bring our column densities up to the observed levels (we limit the metallicity to solar on plausibility grounds).  On the other hand, modifying the temperature (i.e. by lowering it to $3\times 10^5$ K for most radii) while holding everything else fixed succeeds in producing sufficient column densities to match (or even exceed) the observational data, as seen in the right panel of Figure \ref{fig:deduce}.  

This suggests that our MFCS model possesses sufficient quantities of enriched gas at large radii, but that insufficient amounts of cooler gas exist (necessary for O VI).  Note that this procedure is not physically realistic -- we are not free to reduce the gas temperature everywhere (this would decrease the pressure and hence change the gas distribution); however, one solution would be to introduce multi-phase gas, preserving the total pressure but increasing the amount of gas in the $3 \times 10^5$ K temperature range.  This result appears to apply to the other feedback simulations as well, that is, the median temperature is generally too high in our CGM, leading to depleted concentrations of these ions in preference to higher ionization states, but modifying the temperature allows us to match the observations.

Interestingly, the results for the Vanilla run (shown in the left panel of Figure~\ref{fig:deduce}) are quite different -- fine-tuning the temperature in this model cannot bring it in line with observations, since the metallicity falls off far too quickly at $r\sim50$ kpc to allow for the modified temperature to compensate for it, further confirming that feedback and outflows are necessary in order to produce an observationally consistent simulation.

\subsection{Limitations}
\label{limitations}

To better understand the limitations of our models, we look at uncertainties related to photo-ionization and collisional ionization.  To help focus this discussion, we separate our chosen species based on their primary ionization source.  Generally the `high' ions (N V, O VI, O VII) are thought to be associated with collisional ionization, whereas the `low' ions (Mg II, Si II, Si III, Si IV, C IV) can be the result of both collisional and photoionization \citep{2008ApJ...679..194D}.  \citet{2009MNRAS.395.1875O} suggest that many of the weaker O VI absorbers may be due to photoionization, however even they agree that absorbers with equivalent widths greater than 100 m\AA~(most of our observational comparison dataset) are predominantly due to collisions.  With that division made, we can discuss the the ionization sources separately.

\subsubsection{Photoionization}
\label{photoionization}
Our low-energy species are partially ionized by absorption of energetic photons, so any problems in our modeling of photoionization will affect their column density profiles.  Our study calculates photoionization for a metagalactic ionizing background in the optically thin regime, but it does not account for local photoionizing sources, such as energetic stars and galaxies within the volume itself.  Inclusion of these sources will increase the ionization state of the various atoms present, thus potentially increasing (or decreasing) the column density of the ions we study.  Its dominant effect will be felt close to the disk of the galaxy where the bulk of the star formation is occurring and the local radiation field is high.  In a similar study, \citet{Shen:2012ux} included local photoionizing effects with the use of a galactocentric STARBURST99 model\footnote{http://www.stsci.edu/science/starburst99/docs/default.htm} \citep{1999ApJS..123....3L} for a SFR of 20 \msun~yr$^{-1}$ and found that local radiation was only important in the volume $r < 45$ kpc.  Given that our SFRs are typically in the range of 10 \msun~yr$^{-1}$, we predict an even smaller region where local photoionizing sources are important.  However, the discrepancy between observations and the models occurs primarily at large radius, so while the lack of local photoionization may have some effect on our profiles it seems unlikely to explain much of the observed differences in column density profiles for O VI, H I, Mg II and possibly C IV at distance well beyond $r_{\rm inner}$.

Cool gas at high densities will act to shield itself from photoionizing radiation.  In this paper, we have treated all gas in the optically thin regime for absorption of the ionizing background; however, gas with column densities $N_{\rm H I} >10^{17.2}$ cm$^{-2}$ should be partially or completely shielded from radiation more energetic than 13.6 eV \citep[e.g.,][]{2011MNRAS.412L.118F}.  This will have the effect of lowering the ionization state of the various species present, increasing the column density of the lowest ionization species preferentially.  Examination of the H I radial profile reveals that column densities of this magnitude only occur in the near-disk region of the galaxy ($r < r_{\rm inner}$).  Interestingly, our lack of self-shielding may in some cases be balanced by our lack of local photoionizing sources, since they both dominate in the disk and can have counteracting effects depending on the ion.  Either way, neither seem likely to have significant repercussions at large radii, in the region of the greatest observational discrepancies.

Like many simulation groups, we use a Haardt-Madau photoionizing background \citep[][updated to the 2005 version]{Haardt:1996fq}, but there have been some advances in constraining the nature of this background and its sources in the universe since this time.  A few improved approximations of the background have appeared recently \citep[e.g.][]{2009ApJ...703.1416F, 2012ApJ...746..125H}.  Employing a different background has the potential to modify the ionization balance, so as to change the column density of gas in these low-energy ions, although it is unlikely that it will elevate our results by the order of magnitude difference required to align them with observations.

Current work is ongoing to include the aforementioned physics into our simulations.  However, while these effects might make some differences in the simulated profiles, the changes will probably be confined to the near-disk region, failing to bring our models substantially closer to agreement with observations.  

\subsubsection{Collisional Ionization Equilibrium}
The use of Cloudy in calculating the ionization state of our various species assumes collisional ionization equilibrium (CIE), which is an ideal condition requiring $\tau_{\rm cool} >> \tau_{\rm CIE}$ \citep{2003adu..book.....D}.  The CIE assumption holds in high-density, low-temperature regions where the cooling time is short, but it can break down in the low-density regions of the IGM (typically $r > r_{\rm vir}$) \citep{Cen:2006kh}.  \citet{Cen:2006kh} investigated the differences in O VI and O VII absorber strengths with and without the assumption of CIE and found that CIE overpredicts equivalent widths for these two ions by a small factor (less than 50\%). Notably, properly accounting for non-equilibrium effects would decrease our column densities of O VI and O VII, pulling them farther away from the observational sample.

The high-energy species associated with warm and hot gas are primarily ionized due to collisions with ambient electrons.  The rate of these collisions increases with electron density and density of the medium, so that clumpy regions of high density should have a large amount of collisional ionization.  Thus, resolution becomes a key element in order to assure one can resolve clumps sufficiently for proper modeling of their ionization state.  

\subsubsection{Spatial Resolution}
To investigate the impact of resolution and its potential nonlinear effects on the collisional ionization equilibrium in our species, we analyzed several outputs in an attempt to identify the characteristic size of absorbers in various different ionic species.  We found that clumps of material at moderate to high density in our fields came in a variety of sizes, ranging from over 10 kpc down to 425 comoving pc (our spatial resolution limit), but there was no preferred size scale of absorbers.  This continuum of sizes indicates that there may be more gas clumps forming on subgrid scales not resolved in this study.

There is evidence that absorber sizes in reality are below the resolution used in this study.  The size of the background sources, the quasar accretion disks, are typically $\sim0.01$ pc in size \citep{2007ASPC..371...43K}, allowing one to observationally probe very small-scale absorbing structures.  Using multiply-imaged lensed quasars, \citet{1999ApJ...515..500R} find that low ion absorbers can have a length scale $< 20$ pc, a result further confirmed by \citet{2006ApJ...637..648S}.  Photoionization modeling by \citet{2007MNRAS.379.1169S} indicates that high ion absorbers (e.g. C IV) are typically larger but often still quite small, $\sim$ 100 pc.  While not directly CGM related, \citet{2012MNRAS.420.1347F} demonstrate that radiative shocks can develop in AGN outflows generating compact clumps with sizes as small as $\sim$ 0.01 pc.  It is possible such a mechanism might produce similar clumps in galactic wind systems.  Such clumps are potentially significant as a source of multiphase structure and ion absorption below the resolution of our simulations.  It is likely that increasing the spatial resolution of the simulation would better resolve some of these small-scale structures, increasing the overall column density of all low ion absorbers.  This might additionally have an effect on high ion absorbers, as long as their ionization potential had a characteristic temperature less than the virial temperature of the halo.  This effect might reduce the discrepancies between these simulations and observations of the CGM.

\begin{figure*}
\begin{center}
\includegraphics[scale=0.8]{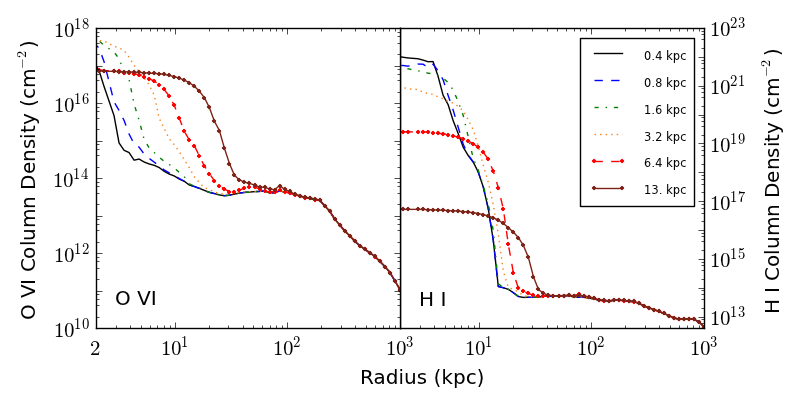}
\end{center}
\caption{Radial profiles of O VI and H I column density as a function of downgraded spatial resolution for a single data output of MFCS at $z=0.2$.}
\label{fig:resolution}
\end{figure*}

To further investigate this issue, we modified outputs from the simulations by resampling them at decreased maximum resolution (0.8, 1.6, 3.2, 6.4 and 13. kpc maximum spatial resolution instead of the 0.4 kpc used in the simulations) prior to post-processing them for the atomic species.  This step is done on all fields (mass-weighted for the temperature and metallicity fields) in order to determine whether or not our atomic post-processing step reaches resolution convergence.  This downsampling will show us where the multiphase substructure exists in the simulation, since smoothing two multiphase structures together (e.g. cold clumps in a hot medium) will result in gas with different ion populations.

Figure \ref{fig:resolution} displays the outcome of this study for O VI and H I column densities of an MFCS output at $z=0.2$.  
Neutral hydrogen, which is very cuspy in the high-resolution simulation (seen as the solid black line) is spread out over an increasingly large region as the spatial resolution diminishes.  The core is particularly sensitive to this effect, as the cold core is artificially mixed with the hotter surrounding medium, it raises the core mean temperature, leading to hydrogen ionization and less H I.  On the other hand, the O VI column density actually increases in this region because the artificial warm component favors O VI over other oxygen ionization states.  In each case, we observe that our change in resolution in post-processing has very little effect on the CGM at large radius (except for a small increase for O VI) suggesting that most of our multiphase substructure exists within $r\sim50$ kpc.  By this measure, we are converging on behavior outside of this radius; however, this result could change if the underlying simulation had higher resolution in the halo, or if it employed a different prescription for generating outflows.  Alternative feedback prescriptions which do not utilize thermal energy deposition (e.g.  radiation pressure from massive stars, kinetic winds from supernovae blastwaves, etc.) may be more successful at generating multiphase outflows by not obliterating the cold, fragile medium.

\subsection{Comparison with Previous Work}
\label{prev_work}
A number of recent studies have investigated the state of the CGM in hydrodynamic galaxy simulations both in the high-redshift and low-redshift universe.  Many such studies focus on neutral hydrogen in the CGM of simulated galaxies at high $z$.
\citet{2011ApJ...737L..37A} post-process an OWLS \citep{2010MNRAS.402.1536S} simulation at $z = 3$ to include a self-shielding correction and the formation of molecular hydrogen, demonstrating that the inclusion of these features causes a flattening and steepening of the H I column density distribution function at $N_{\rm HI} \sim 10^{18}$ cm$^{-2}$ and $N_{\rm HI} \sim 10^{21}$ cm$^{-2}$ respectively.\ 
\citet{2011MNRAS.412L.118F} examine an SPH simulation of a galaxy at $z=2-4$ by ray-tracing the ionizing background to correct for self-shielding, discovering low covering factors of high-density H I in the CGM consistent with those we produce in our models.
\citet{2011MNRAS.416.1723B} investigate how modifying the intensity of galactic winds affects the H I distribution and subsequent Lyman-$\alpha$ emission from such systems at $z = 3$.  Their momentum-driven wind feedback prescription yields H I distributions similar to our own MFCS run at that redshift.  

Other researchers have explored using metal ions as possible probes of the CGM by post-processing their simulations to include additional species, as we do in the current study.
\citet{Kacprzak:2010ds} added a Mg II field to an AMR galaxy simulation at $z=1$ (similar to our MF run), finding a Mg II-absorber size distribution and Mg II $r_{\rm inner} =  20$ kpc consistent with our own work. 
\citet{2011MNRAS.418.1796F} examined outputs at $z\sim2-3$ from an AMR simulation like our MF model, processing them to include self-shielding due to dust, radiative transfer from local ionizing sources, and H I and Si II fields.  In agreement with our results, they determined that the concentration of Si II absorbers is an order of magnitude lower than observed values, suggesting they too may lack significant multiphase structures in the CGM.
A suite of simulations modifying the feedback prescription to include AGN and wind-driven feedback was performed by \citet{2011MNRAS.411..826T}, demonstrating that C IV statistics were considerably more sensitive to feedback implementation than H I in the intergalactic medium (IGM) at $z=2-3$.  While their conclusion agrees with our results in the outer parts of our halos ($r > 200$ kpc), we find that both H I and C IV are sensitive to feedback prescription in the CGM ($20 < r < 200$ kpc) as exhibited in Figure \ref{fig:radprof2}.
\citet{Goerdt:2012vk} post-process an AMR simulation of $z\sim2.5$ galaxy output to include a self-shielding correction and H I, O I, C II, C IV, Si II, Si IV, Mg II, and Fe II, observing the distribution of these metals versus galactocentric radius using a feedback model similar to our MF simulation.  In general our results at that redshift align well with their study, except our C IV column densities are an order of magnitude greater than theirs, which might be a due to their self-shielding correction suppressing higher levels of ionization in the CGM.  
Lastly, \citet{Shen:2012ux} analyze a $z=2.8$ output from the Eris Simulation \citep{2011ApJ...742...76G} akin to the MFCS model, including a spherically-symmetric correction for local photoionizing sources and adding H I, C II, Si II, Si IV, C IV, and O VI in post-processing.  They show the radial dependence of O VI column density, generally agreeing (ours is flatter) with our MFCS results at $z\sim3$ (see Figure \ref{fig:redshift}).  
  
At low redshifts, there have only been a few studies probing the nature of the CGM in numerical simulations.  \citet{VanDeVoort:2011wk} demonstrated that inflowing, cold gas lacks significant metals in the CGM, consistent with our vanilla run presented in Figure \ref{fig:radprof1}. 
Turning to H I, \citet{2012ApJ...749..181F} looked at a $z=0$ galaxy (similar to our MF model) and examined the source of H I feeding its CGM.  They showed H I column densities and covering factors matching our own study, exhibiting a steep falloff of $N_{\rm HI}$ at $r\sim10$ kpc but an extended floor of material beyond $r_{\rm vir}$.  \citet{Ford:2012tr} analyzed a series of SPH simulation outputs at $z=0.25$ to investigate the column density of various metal lines in proximity to a large samples of galaxies of different characteristic mass.  The results for their fiducial feedback model (utilizing explicit momentum-driven winds) on 10$^{12}$ \msun~halos match well against our MFCS run (as seen in their Figures 2, 11, and 12), and the apparent discrepancies in Figures 9 and 10 appear to be due to the additional normalization they employ for dealing with multiple halos.
\citet{Stinson:2011tp} produced a suite of SPH simulations run to $z=0$ including two feedback intensities.  Their ``low'' feedback run was similar to our LF run but included cooling suppression, whereas their ``high'' feedback run (intensified in order to integrate the effects of radiation from massive, young stars) was $\sim3\times$ stronger than our HF run while still including a form of cooling suppression.  The resulting outputs were processed to include O VI data, and then radial profiles were produced for both H I and O VI.  The column density profiles for the ``low'' feedback run are consistent with our LF simulation; however, the profiles of their ``high'' feedback run have column densities exceeding any of our simulations.  These profiles align well with the observational datasets in H I and O VI, suggesting that their success may be a result of using extremely high feedback intensities (i.e. $10^{51}$ ergs released per 7.5 \msun~of stars formed).  We see a possible trend in our models, potentially indicating that intensifying feedback could increase column densities of various species in the CGM region (notably O VI) but this prediction is not clear cut.  

There exists an additional physical effect that might account for some of the discrepancy between the \citet{Stinson:2011tp} work and our own.  Much of the gas in these simulated galaxies resides at the virial temperature of the halo $T_{\rm vir} \propto M_{\rm halo} / r_{\rm halo}$.  Our simulated halo is $M_{200} = 1.1 \times 10^{12}$ \msun\ whereas the halo used in their high feedback run has a mass $M_{\rm vir} = 7 \times 10^{11}$ \msun.  This equates to roughly to a 35\% decrease in the virial temperature of their halo relative to ours.  As already noted, the O VI transition has a relatively small temperature window, centered around $T \sim 3 \times 10^5$ K.  A modest shift in the ambient temperature of a halo could yield a significantly increase in the amount of O VI present, potentially explaining some of the disagreement in the results of these two studies.

\section{Conclusions}
\label{conclusions}

We perform the first AMR study to probe the CGM at low redshifts by comparing simulations against a large host of observational probes and the first study to look at the redshift evolution of the simulated CGM.  We carry out cosmological simulations of a Milky-Way-massed galactic halo using a variety of stellar feedback prescriptions, post-processing the data to compute the column density distributions of a wide range of atomic species, spanning an ionization energy range from 13.6 eV to 739 eV to effectively probe different phases of the CGM.   We median-average these distributions over many orientations and times (from $z=0.1$ to $z=0.5$) both in order to make a cleaner comparison with data, and to decrease the impact of satellites on our resulting radial profiles.  Lastly, we compare these results against a number of different low-z observational datasets, correlating absorber strength and impact parameter to the nearest galaxy galaxies.  Our main results are as follows.

\begin{itemize}
\item It is difficult for any of our feedback models to reproduce the column density distribution of some of the observed absorbers -- in particular, we cannot match observations of Mg II beyond $r > 50$ kpc, and O VI at any radii.  On the other hand, we find that the Mg II distribution at smaller radii ($r < 50$ kpc) can be reproduced in some of the models.  In addition, the predicted O VII absorption (which probes hot gas) in simulations including moderate or strong feedback is in agreement with current observations.

\item We find that some form of feedback is required to get significant amounts of metals into the CGM beyond $\sim50$ kpc, but the total amount of gas in this region of the CGM is not strongly affected by feedback.  In models which include some minimal amount of feedback, we find sufficient metals in our simulated CGM to reproduce observations, but a lack of multi-phase gas means that not enough of the gas has sufficiently low temperatures to produce the observed ions.  This is clearest in the case of O VI, where the observations generally do not match any of our models because of this mismatched phase (i.e. temperature).

\item Turning thermal feedback up to very high values (even without cooling suppression) is successful at producing a galaxy with a flat rotation curve; however, it still fails to match the observational datasets from QSO absorption line studies (notably O VI) demonstrating that QSO observations provide additional constraints on subgrid models beyond traditional metrics of success.

\item There is significant redshift evolution of the column density distribution for our atomic species.  In general, the column densities for most species drop by almost an order of magnitude across all radii in evolving from $z \sim 3$ to $z = 0$.

\end{itemize}

Generally, we find that QSO observations place tight constraints on feedback models.  Perhaps not surprisingly, the CGM which acts as both stellar fuel and the medium into which supernovae explode, contains a great deal of information regarding the specific nature of star formation and stellar feedback.  Probing this medium through QSO absorption line studies can tap into this information and reveal clear ways of differentiating between subgrid models.  Significantly sampled observations in any of H I, Si II, Si III, Si IV, or C IV identifying $r_{\rm inner}$, the transition radius where column densities plummet in the CGM for these low-ionization ions, should potentially favor a specific type and intensity of stellar feedback.  Furthermore, even subgrid models which can meet other criteria for success (e.g. flat rotation curve, matching the Kennicutt-Schmidt Law) need not match QSO absorption line statistics, as is demonstrated in the case of our HF run producing a flat rotation curve but failing to reproduce the the Mg II, H I and O VI observational datasets.  Thus, comparisons against QSO absorption lines place a constraint on galaxy simulations orthogonal to standard comparison modes.

Future work on this topic includes addressing some of the limitations considered in Section \ref{limitations}.  As was discussed, it appears that we lack significant multiphase substructure in the halos of our galaxies, which is likely responsible for our inability to produce sufficient column densities of certain atomic species required to match observations (e.g. H I, Mg II, O VI).  The use of subgrid models which deposit stellar feedback explosions as thermal energy may be responsible for generating unrealistic hot outflows.  Relying solely on ram pressure from hot winds to distribute feedback energy could result in the spurious destruction of any existing cold halo gas and suppress the creation of a multiphase medium in the CGM.  Recent work on alternative feedback prescriptions may provide a solution.  For instance, \citet{2011ApJ...735...66M} emphasize that radiation pressure from energetic massive stars on the surrounding dust-entrained gas may provide a way to get cool gas out to radii where it is commonly observed.  There have been a few promising demonstrations of radiative feedback \citep[e.g.][]{2011ApJ...738...54K, Kim:2012va, 2012MNRAS.421.3522H, 2012MNRAS.427..311W}.  Likewise, subgrid models of feedback where supernova energy is deposited as kinetic winds \citep[e.g.][]{Springel:2003eg, Agertz:2012wq} or as cosmic rays \citep[e.g.][]{Miniati:2001bz, Jubelgas:2008je, 2012MNRAS.421.3375V} might also avoid destroying a fragile multiphase medium.   Additionally, steps to ensure higher resolution in the galactic halo may result in smaller clumps forming there; however, one potentially must modify refinement criteria to achieve substantial resolution in the low-density halo.

\section{Acknowledgements}

CBH and GLB acknowledge support from NSF grants AST-0547823, AST-0908390, and AST-1008134, as well as computational resources from NSF XSEDE and Columbia's hotfoot cluster.  BDS was supported in part by NASA through grant \#NNX09AD80G and by the NSF through AST grant 0908819.  MJT acknowledges support by NSF CI TraCS
fellowship award OCI-1048505.  We also thank the \emph{Enzo} and \emph{yt} communities for their helpful discussion and problem-solving on various aspects of the production and analysis of this work.  We benefitted from discussions on this topic with Hsiao-Wen Chen, Arlin Crotts, Claude-Andr\'{e} Faucher-Gigu\`{e}re, Amanda Ford, M. Ryan Joung, Jeremiah Ostriker, Mary Putman, David Schiminovich, and Art Wolfe.

\bibliography{paper2.bib}

\begin{thebibliography}{}

\bibitem[\protect\citeauthoryear{Abadi, Navarro, Steinmetz \& Eke}{Abadi
  et~al.}{2003}]{Abadi:2003bs}
Abadi M.~G.,  Navarro J.~F.,  Steinmetz M.,    Eke V.~R.,  2003, The
  Astrophysical Journal, 591, 499

\bibitem[\protect\citeauthoryear{Abel, Anninos, Zhang \& Norman}{Abel
  et~al.}{1997}]{Abel:1997jz}
Abel T.,  Anninos P.,  Zhang Y.,    Norman M.~L.,  1997, New Astronomy, 2, 181

\bibitem[\protect\citeauthoryear{Agertz, Kravtsov, Leitner \& Gnedin}{Agertz
  et~al.}{2012}]{Agertz:2012wq}
Agertz O.,  Kravtsov A.~V.,  Leitner S.~N.,    Gnedin N.~Y.,  2012, arXiv,
  astro-ph.CO

\bibitem[\protect\citeauthoryear{Agertz, Teyssier \& Moore}{Agertz
  et~al.}{2011}]{2011MNRAS.410.1391A}
Agertz O.,  Teyssier R.,    Moore B.,  2011, Monthly Notices of the Royal
  Astronomical Society, 410, 1391

\bibitem[\protect\citeauthoryear{Altay, Theuns, Schaye, Crighton \&
  Dalla~Vecchia}{Altay et~al.}{2011}]{2011ApJ...737L..37A}
Altay G.,  Theuns T.,  Schaye J.,  Crighton N. H.~M.,    Dalla~Vecchia C.,
  2011, The Astrophysical Journal Letters, 737, L37

\bibitem[\protect\citeauthoryear{Anninos, Zhang, Abel \& Norman}{Anninos
  et~al.}{1997}]{Anninos:1997in}
Anninos P.,  Zhang Y.,  Abel T.,    Norman M.~L.,  1997, New Astronomy, 2, 209

\bibitem[\protect\citeauthoryear{Barnes, Haehnelt, Tescari \& Viel}{Barnes
  et~al.}{2011}]{2011MNRAS.416.1723B}
Barnes L.~A.,  Haehnelt M.~G.,  Tescari E.,    Viel M.,  2011, Monthly Notices
  of the Royal Astronomical Society, 416, 1723

\bibitem[\protect\citeauthoryear{Behroozi, Conroy \& Wechsler}{Behroozi
  et~al.}{2010}]{Behroozi:2010ja}
Behroozi P.~S.,  Conroy C.,    Wechsler R.~H.,  2010, The Astrophysical
  Journal, 717, 379

\bibitem[\protect\citeauthoryear{Binney}{Binney}{1977}]{1977ApJ...215..483B}
Binney J.,  1977, Astrophysical Journal, 215, 483

\bibitem[\protect\citeauthoryear{Birnboim \& Dekel}{Birnboim \&
  Dekel}{2003}]{2003MNRAS.345..349B}
Birnboim Y.,  Dekel A.,  2003, Monthly Notices of the Royal Astronomical
  Society, 345, 349

\bibitem[\protect\citeauthoryear{Booth \& Schaye}{Booth \&
  Schaye}{2009}]{2009MNRAS.398...53B}
Booth C.~M.,  Schaye J.,  2009, Monthly Notices of the Royal Astronomical
  Society, 398, 53

\bibitem[\protect\citeauthoryear{Bryan \& Norman}{Bryan \&
  Norman}{1997}]{Bryan:1997wy}
Bryan G.~L.,  Norman M.~L.,  1997, arXiv, astro-ph, 363

\bibitem[\protect\citeauthoryear{Cen \& Fang}{Cen \& Fang}{2006}]{Cen:2006kh}
Cen R.,  Fang T.,  2006, The Astrophysical Journal, 650, 573

\bibitem[\protect\citeauthoryear{Cen \& Ostriker}{Cen \&
  Ostriker}{1992}]{Cen:1992ju}
Cen R.,  Ostriker J.~P.,  1992, The Astrophysical Journal, 399, L113

\bibitem[\protect\citeauthoryear{Chen, Helsby, Gauthier, Shectman, Thompson \&
  Tinker}{Chen et~al.}{2010}]{Chen:2010ir}
Chen H.-W.,  Helsby J.~E.,  Gauthier J.-R.,  Shectman S.~A.,  Thompson I.~B.,
   Tinker J.~L.,  2010, The Astrophysical Journal, 714, 1521

\bibitem[\protect\citeauthoryear{Chen, Lanzetta, Webb \& Barcons}{Chen
  et~al.}{2001}]{Chen:2001hj}
Chen H.-W.,  Lanzetta K.~M.,  Webb J.~K.,    Barcons X.,  2001, The
  Astrophysical Journal, 559, 654

\bibitem[\protect\citeauthoryear{Chen \& Tinker}{Chen \&
  Tinker}{2008}]{Chen:2008cd}
Chen H.-W.,  Tinker J.~L.,  2008, The Astrophysical Journal, 687, 745

\bibitem[\protect\citeauthoryear{Col{\'\i}n, Avila-Reese, V{\'a}zquez-Semadeni,
  Valenzuela \& Ceverino}{Col{\'\i}n et~al.}{2010}]{Colin:2010ee}
Col{\'\i}n P.,  Avila-Reese V.,  V{\'a}zquez-Semadeni E.,  Valenzuela O.,
  Ceverino D.,  2010, The Astrophysical Journal, 713, 535

\bibitem[\protect\citeauthoryear{Danforth \& Shull}{Danforth \&
  Shull}{2008}]{2008ApJ...679..194D}
Danforth C.~W.,  Shull J.~M.,  2008, The Astrophysical Journal, 679, 194

\bibitem[\protect\citeauthoryear{Dav{\'e}, Finlator \& Oppenheimer}{Dav{\'e}
  et~al.}{2012}]{2012MNRAS.421...98D}
Dav{\'e} R.,  Finlator K.,    Oppenheimer B.~D.,  2012, Monthly Notices of the
  Royal Astronomical Society, 421, 98

\bibitem[\protect\citeauthoryear{Dopita \& Sutherland}{Dopita \&
  Sutherland}{2003}]{2003adu..book.....D}
Dopita M.~A.,  Sutherland R.~S.,  2003, {Astrophysics of the Diffuse Universe}.
Astronomy and Astrophysics Library, Springer, Berlin, New York

\bibitem[\protect\citeauthoryear{Draine}{Draine}{2011}]{Draine:2011tr}
Draine B.~T.,  2011, Physics of the Interstellar and Intergalactic Medium by
  Bruce T. Draine. Princeton University Press, 2011. ISBN: 978-0-691-12214-4,
  -1

\bibitem[\protect\citeauthoryear{Efstathiou, Davis, White \& Frenk}{Efstathiou
  et~al.}{1985}]{1985ApJS...57..241E}
Efstathiou G.,  Davis M.,  White S. D.~M.,    Frenk C.~S.,  1985, Astrophysical
  Journal Supplement Series (ISSN 0067-0049), 57, 241

\bibitem[\protect\citeauthoryear{Faucher-Gigu{\`e}re \& Kere{\v
  s}}{Faucher-Gigu{\`e}re \& Kere{\v s}}{2011}]{2011MNRAS.412L.118F}
Faucher-Gigu{\`e}re C.-A.,  Kere{\v s} D.,  2011, Monthly Notices of the Royal
  Astronomical Society: Letters, 412, L118

\bibitem[\protect\citeauthoryear{Faucher-Gigu{\`e}re, Kere{\v s} \&
  Ma}{Faucher-Gigu{\`e}re et~al.}{2011}]{2011MNRAS.417.2982F}
Faucher-Gigu{\`e}re C.-A.,  Kere{\v s} D.,    Ma C.-P.,  2011, Monthly Notices
  of the Royal Astronomical Society, 417, 2982

\bibitem[\protect\citeauthoryear{Faucher-Gigu{\`e}re, Lidz, Zaldarriaga \&
  Hernquist}{Faucher-Gigu{\`e}re et~al.}{2009}]{2009ApJ...703.1416F}
Faucher-Gigu{\`e}re C.-A.,  Lidz A.,  Zaldarriaga M.,    Hernquist L.,  2009,
  The Astrophysical Journal, 703, 1416

\bibitem[\protect\citeauthoryear{Faucher-Gigu{\`e}re, Quataert \&
  Murray}{Faucher-Gigu{\`e}re et~al.}{2012}]{2012MNRAS.420.1347F}
Faucher-Gigu{\`e}re C.-A.,  Quataert E.,    Murray N.,  2012, Monthly Notices
  of the Royal Astronomical Society, 420, 1347

\bibitem[\protect\citeauthoryear{Ferland, Korista, Verner, Ferguson, Kingdon \&
  Verner}{Ferland et~al.}{1998}]{1998PASP..110..761F}
Ferland G.~J.,  Korista K.~T.,  Verner D.~A.,  Ferguson J.~W.,  Kingdon J.~B.,
    Verner E.~M.,  1998, The Publications of the Astronomical Society of the
  Pacific, 110, 761

\bibitem[\protect\citeauthoryear{Fern{\'a}ndez, Joung \& Putman}{Fern{\'a}ndez
  et~al.}{2012}]{2012ApJ...749..181F}
Fern{\'a}ndez X.,  Joung M.~R.,    Putman M.~E.,  2012, The Astrophysical
  Journal, 749, 181

\bibitem[\protect\citeauthoryear{Ford, Oppenheimer, Dav{\'e}, Katz, Kollmeier
  \& Weinberg}{Ford et~al.}{2012}]{Ford:2012tr}
Ford A.~B.,  Oppenheimer B.~D.,  Dav{\'e} R.,  Katz N.,  Kollmeier J.~A.,
  Weinberg D.~H.,  2012, arXiv, astro-ph.CO

\bibitem[\protect\citeauthoryear{Fumagalli, Prochaska, Kasen, Dekel, Ceverino
  \& Primack}{Fumagalli et~al.}{2011}]{2011MNRAS.418.1796F}
Fumagalli M.,  Prochaska J.~X.,  Kasen D.,  Dekel A.,  Ceverino D.,    Primack
  J.~R.,  2011, Monthly Notices of the Royal Astronomical Society, 418, 1796

\bibitem[\protect\citeauthoryear{Gerritsen}{Gerritsen}{1997}]{Gerritsen:1997tf}
Gerritsen J.,  1997, Ph.D. thesis, Groningen University, the Netherlands,
  (1997)

\bibitem[\protect\citeauthoryear{Goerdt, Dekel, Sternberg, Gnat \&
  Ceverino}{Goerdt et~al.}{2012}]{Goerdt:2012vk}
Goerdt T.,  Dekel A.,  Sternberg A.,  Gnat O.,    Ceverino D.,  2012, arXiv,
  astro-ph.CO

\bibitem[\protect\citeauthoryear{Governato, Willman, Mayer, Brooks, Stinson,
  Valenzuela, Wadsley \& Quinn}{Governato et~al.}{2007}]{Governato:2007dq}
Governato F.,  Willman B.,  Mayer L.,  Brooks A.,  Stinson G.,  Valenzuela O.,
  Wadsley J.,    Quinn T.,  2007, Monthly Notices of the Royal Astronomical
  Society, 374, 1479

\bibitem[\protect\citeauthoryear{Governato, Zolotov, Pontzen, Christensen, Oh,
  Brooks, Quinn, Shen \& Wadsley}{Governato et~al.}{2012}]{2012MNRAS.422.1231G}
Governato F.,  Zolotov A.,  Pontzen A.,  Christensen C.,  Oh S.~H.,  Brooks
  A.~M.,  Quinn T.,  Shen S.,    Wadsley J.,  2012, Monthly Notices of the
  Royal Astronomical Society, 422, 1231

\bibitem[\protect\citeauthoryear{Guedes, Callegari, Madau \& Mayer}{Guedes
  et~al.}{2011}]{2011ApJ...742...76G}
Guedes J.,  Callegari S.,  Madau P.,    Mayer L.,  2011, The Astrophysical
  Journal, 742, 76

\bibitem[\protect\citeauthoryear{Haardt \& Madau}{Haardt \&
  Madau}{1996}]{Haardt:1996fq}
Haardt F.,  Madau P.,  1996, The Astrophysical Journal, 461, 20

\bibitem[\protect\citeauthoryear{Haardt \& Madau}{Haardt \&
  Madau}{2012}]{2012ApJ...746..125H}
Haardt F.,  Madau P.,  2012, The Astrophysical Journal, 746, 125

\bibitem[\protect\citeauthoryear{Hopkins, Quataert \& Murray}{Hopkins
  et~al.}{2012}]{2012MNRAS.421.3522H}
Hopkins P.~F.,  Quataert E.,    Murray N.,  2012, Monthly Notices of the Royal
  Astronomical Society, 421, 3522

\bibitem[\protect\citeauthoryear{Hummels \& Bryan}{Hummels \&
  Bryan}{2012}]{2012ApJ...749..140H}
Hummels C.~B.,  Bryan G.~L.,  2012, The Astrophysical Journal, 749, 140

\bibitem[\protect\citeauthoryear{Joung, Putman, Bryan, Fern{\'a}ndez \&
  Peek}{Joung et~al.}{2012}]{Joung:2012wj}
Joung M.~R.,  Putman M.~E.,  Bryan G.~L.,  Fern{\'a}ndez X.,    Peek J.,  2012,
  arXiv, astro-ph.CO

\bibitem[\protect\citeauthoryear{Jubelgas, Springel, En{\ss}lin \&
  Pfrommer}{Jubelgas et~al.}{2008}]{Jubelgas:2008je}
Jubelgas M.,  Springel V.,  En{\ss}lin T.,    Pfrommer C.,  2008, Astronomy and
  Astrophysics, 481, 33

\bibitem[\protect\citeauthoryear{Kacprzak, Churchill, Ceverino, Steidel, Klypin
  \& Murphy}{Kacprzak et~al.}{2010}]{Kacprzak:2010ds}
Kacprzak G.~G.,  Churchill C.~W.,  Ceverino D.,  Steidel C.~C.,  Klypin A.,
  Murphy M.~T.,  2010, The Astrophysical Journal, 711, 533

\bibitem[\protect\citeauthoryear{Katz}{Katz}{1992}]{Katz:1992gu}
Katz N.,  1992, The Astrophysical Journal, 391, 502

\bibitem[\protect\citeauthoryear{Kere{\v s}, Katz, Fardal, Dav{\'e} \&
  Weinberg}{Kere{\v s} et~al.}{2009}]{Keres:2009bq}
Kere{\v s} D.,  Katz N.,  Fardal M.,  Dav{\'e} R.,    Weinberg D.~H.,  2009,
  Monthly Notices of the Royal Astronomical Society, 395, 160

\bibitem[\protect\citeauthoryear{Kere{\v s}, Katz, Weinberg \&
  Dav{\'e}}{Kere{\v s} et~al.}{2005}]{Keres:2005gb}
Kere{\v s} D.,  Katz N.,  Weinberg D.~H.,    Dav{\'e} R.,  2005, Monthly
  Notices of the Royal Astronomical Society, 363, 2

\bibitem[\protect\citeauthoryear{Kim, Krumholz, Wise, Turk, Goldbaum \&
  Abel}{Kim et~al.}{2012}]{Kim:2012va}
Kim J.-h.,  Krumholz M.~R.,  Wise J.~H.,  Turk M.~J.,  Goldbaum N.~J.,    Abel
  T.,  2012, arXiv, astro-ph.GA

\bibitem[\protect\citeauthoryear{Kim, Wise, Alvarez \& Abel}{Kim
  et~al.}{2011}]{2011ApJ...738...54K}
Kim J.-h.,  Wise J.~H.,  Alvarez M.~A.,    Abel T.,  2011, The Astrophysical
  Journal, 738, 54

\bibitem[\protect\citeauthoryear{Kimm, Slyz, Devriendt \& Pichon}{Kimm
  et~al.}{2011}]{2011MNRAS.413L..51K}
Kimm T.,  Slyz A.,  Devriendt J.,    Pichon C.,  2011, Monthly Notices of the
  Royal Astronomical Society: Letters, 413, L51

\bibitem[\protect\citeauthoryear{Kochanek, Dai, Morgan, Morgan \&
  Poindexter}{Kochanek et~al.}{2007}]{2007ASPC..371...43K}
Kochanek C.~S.,  Dai X.,  Morgan C.,  Morgan N.,    Poindexter S. C.~G.,  2007,
  Statistical Challenges in Modern Astronomy IV ASP Conference Series, 371, 43

\bibitem[\protect\citeauthoryear{Komatsu, Dunkley, Nolta, Bennett, Gold,
  Hinshaw, Jarosik, Larson, Limon, Page, Spergel, Halpern, Hill, Kogut, Meyer,
  Tucker, Weiland, Wollack \& Wright}{Komatsu et~al.}{2009}]{Komatsu:2009ex}
Komatsu E.,  Dunkley J.,  Nolta M.~R.,  Bennett C.~L.,  Gold B.,  Hinshaw G.,
  Jarosik N.,  Larson D.,  Limon M.,  Page L.,  Spergel D.~N.,  Halpern M.,
  Hill R.~S.,  Kogut A.,  Meyer S.~S.,  Tucker G.~S.,  Weiland J.~L.,  Wollack
  E.,    Wright E.~L.,  2009, The Astrophysical Journal Supplement, 180, 330

\bibitem[\protect\citeauthoryear{Leitherer, Schaerer, Goldader,
  Gonz{\'a}lez~Delgado, Robert, Kune, de Mello, Devost \& Heckman}{Leitherer
  et~al.}{1999}]{1999ApJS..123....3L}
Leitherer C.,  Schaerer D.,  Goldader J.~D.,  Gonz{\'a}lez~Delgado R.~M.,
  Robert C.,  Kune D.~F.,  de Mello D.~F.,  Devost D.,    Heckman T.~M.,  1999,
  The Astrophysical Journal, 123, 3

\bibitem[\protect\citeauthoryear{Machacek, Bryan \& Abel}{Machacek
  et~al.}{2001}]{Machacek:2001fq}
Machacek M.~E.,  Bryan G.~L.,    Abel T.,  2001, The Astrophysical Journal,
  548, 509

\bibitem[\protect\citeauthoryear{Mandelbaum, Seljak, Kauffmann, Hirata \&
  Brinkmann}{Mandelbaum et~al.}{2006}]{2006MNRAS.368..715M}
Mandelbaum R.,  Seljak U.,  Kauffmann G.,  Hirata C.~M.,    Brinkmann J.,
  2006, Monthly Notices of the Royal Astronomical Society, 368, 715

\bibitem[\protect\citeauthoryear{Miniati}{Miniati}{2001}]{Miniati:2001bz}
Miniati F.,  2001, Computer Physics Communications, 141, 17

\bibitem[\protect\citeauthoryear{Murray, M{\'e}nard \& Thompson}{Murray
  et~al.}{2011}]{2011ApJ...735...66M}
Murray N.,  M{\'e}nard B.,    Thompson T.~A.,  2011, The Astrophysical Journal,
  735, 66

\bibitem[\protect\citeauthoryear{Ocvirk, Pichon \& Teyssier}{Ocvirk
  et~al.}{2008}]{2008MNRAS.390.1326O}
Ocvirk P.,  Pichon C.,    Teyssier R.,  2008, Monthly Notices of the Royal
  Astronomical Society, 390, 1326

\bibitem[\protect\citeauthoryear{Oppenheimer \& Dav{\'e}}{Oppenheimer \&
  Dav{\'e}}{2006}]{Oppenheimer:2006eq}
Oppenheimer B.~D.,  Dav{\'e} R.,  2006, Monthly Notices of the Royal
  Astronomical Society, 373, 1265

\bibitem[\protect\citeauthoryear{Oppenheimer \& Dav{\'e}}{Oppenheimer \&
  Dav{\'e}}{2008}]{Oppenheimer:2008bu}
Oppenheimer B.~D.,  Dav{\'e} R.,  2008, Monthly Notices of the Royal
  Astronomical Society, 387, 577

\bibitem[\protect\citeauthoryear{Oppenheimer \& Dav{\'e}}{Oppenheimer \&
  Dav{\'e}}{2009}]{2009MNRAS.395.1875O}
Oppenheimer B.~D.,  Dav{\'e} R.,  2009, Monthly Notices of the Royal
  Astronomical Society, 395, 1875

\bibitem[\protect\citeauthoryear{O'Shea, Bryan, Bordner, Norman, Abel, Harkness
  \& Kritsuk}{O'Shea et~al.}{2004}]{OShea:2004tr}
O'Shea B.~W.,  Bryan G.,  Bordner J.,  Norman M.~L.,  Abel T.,  Harkness R.,
  Kritsuk A.,  2004, arXiv, astro-ph

\bibitem[\protect\citeauthoryear{Penton, Stocke \& Shull}{Penton
  et~al.}{2004}]{2004ApJS..152...29P}
Penton S.~V.,  Stocke J.~T.,    Shull J.~M.,  2004, The Astrophysical Journal,
  152, 29

\bibitem[\protect\citeauthoryear{Piontek \& Steinmetz}{Piontek \&
  Steinmetz}{2011}]{2011MNRAS.410.2625P}
Piontek F.,  Steinmetz M.,  2011, Monthly Notices of the Royal Astronomical
  Society, 410, 2625

\bibitem[\protect\citeauthoryear{Press, Flannery, Teukolsky \&
  Vetterling}{Press et~al.}{1992}]{num_recipes}
Press W.~H.,  Flannery B.~P.,  Teukolsky S.~A.,    Vetterling W.~T.,  1992,
  {Numerical Recipes in FORTRAN: The Art of Scientific Computing}, 2nd edition
  edn.
Cambridge University Press, Cambridge, England

\bibitem[\protect\citeauthoryear{Prochaska, Weiner, Chen, Mulchaey \&
  Cooksey}{Prochaska et~al.}{2011}]{2011ApJ...740...91P}
Prochaska J.~X.,  Weiner B.,  Chen H.-W.,  Mulchaey J.,    Cooksey K.,  2011,
  The Astrophysical Journal, 740, 91

\bibitem[\protect\citeauthoryear{Putman, Peek \& Joung}{Putman
  et~al.}{2012}]{Putman:2012vp}
Putman M.,  Peek J.,    Joung M.,  2012, Annual Reviews in Astrophysics, pp
  1--49

\bibitem[\protect\citeauthoryear{Rauch, Sargent \& Barlow}{Rauch
  et~al.}{1999}]{1999ApJ...515..500R}
Rauch M.,  Sargent W. L.~W.,    Barlow T.~A.,  1999, The Astrophysical Journal,
  515, 500

\bibitem[\protect\citeauthoryear{Ribaudo, Lehner \& Howk}{Ribaudo
  et~al.}{2011}]{2011ApJ...736...42R}
Ribaudo J.,  Lehner N.,    Howk J.~C.,  2011, The Astrophysical Journal, 736,
  42

\bibitem[\protect\citeauthoryear{Scannapieco, Tissera, White \&
  Springel}{Scannapieco et~al.}{2006}]{Scannapieco:2006ij}
Scannapieco C.,  Tissera P.~B.,  White S. D.~M.,    Springel V.,  2006, Monthly
  Notices of the Royal Astronomical Society, 371, 1125

\bibitem[\protect\citeauthoryear{Schaye, Carswell \& Kim}{Schaye
  et~al.}{2007}]{2007MNRAS.379.1169S}
Schaye J.,  Carswell R.~F.,    Kim T.-S.,  2007, Monthly Notices of the Royal
  Astronomical Society, 379, 1169

\bibitem[\protect\citeauthoryear{Schaye, Dalla~Vecchia, Booth, Wiersma, Theuns,
  Haas, Bertone, Duffy, McCarthy \& Van De~Voort}{Schaye
  et~al.}{2010}]{2010MNRAS.402.1536S}
Schaye J.,  Dalla~Vecchia C.,  Booth C.~M.,  Wiersma R. P.~C.,  Theuns T.,
  Haas M.~R.,  Bertone S.,  Duffy A.~R.,  McCarthy I.~G.,    Van De~Voort F.,
  2010, Monthly Notices of the Royal Astronomical Society, 402, 1536

\bibitem[\protect\citeauthoryear{Sharma \& Nath}{Sharma \&
  Nath}{2012}]{2012ApJ...750...55S}
Sharma M.,  Nath B.~B.,  2012, The Astrophysical Journal, 750, 55

\bibitem[\protect\citeauthoryear{Shen, Madau, Guedes, Mayer \& Prochaska}{Shen
  et~al.}{2012}]{Shen:2012ux}
Shen S.,  Madau P.,  Guedes J.,  Mayer L.,    Prochaska J.~X.,  2012, arXiv,
  astro-ph.CO

\bibitem[\protect\citeauthoryear{Simcoe, Sargent \& Rauch}{Simcoe
  et~al.}{2004}]{2004ApJ...606...92S}
Simcoe R.~A.,  Sargent W. L.~W.,    Rauch M.,  2004, The Astrophysical Journal,
  606, 92

\bibitem[\protect\citeauthoryear{Simcoe, Sargent, Rauch \& Becker}{Simcoe
  et~al.}{2006}]{2006ApJ...637..648S}
Simcoe R.~A.,  Sargent W. L.~W.,  Rauch M.,    Becker G.,  2006, The
  Astrophysical Journal, 637, 648

\bibitem[\protect\citeauthoryear{Smith, Sigurdsson \& Abel}{Smith
  et~al.}{2008}]{2008MNRAS.385.1443S}
Smith B.,  Sigurdsson S.,    Abel T.,  2008, Monthly Notices of the Royal
  Astronomical Society, 385, 1443

\bibitem[\protect\citeauthoryear{Smith, Hallman, Shull \& O'Shea}{Smith
  et~al.}{2011}]{2011ApJ...731....6S}
Smith B.~D.,  Hallman E.~J.,  Shull J.~M.,    O'Shea B.~W.,  2011, The
  Astrophysical Journal, 731, 6

\bibitem[\protect\citeauthoryear{Sommer-Larsen, G{\"o}tz \&
  Portinari}{Sommer-Larsen et~al.}{2003}]{SommerLarsen:2003eq}
Sommer-Larsen J.,  G{\"o}tz M.,    Portinari L.,  2003, The Astrophysical
  Journal, 596, 47

\bibitem[\protect\citeauthoryear{Springel \& Hernquist}{Springel \&
  Hernquist}{2003}]{Springel:2003eg}
Springel V.,  Hernquist L.,  2003, Monthly Notice of the Royal Astronomical
  Society, 339, 289

\bibitem[\protect\citeauthoryear{Steidel, Erb, Shapley, Pettini, Reddy,
  Bogosavljevi{\'c}, Rudie \& Rakic}{Steidel
  et~al.}{2010}]{2010ApJ...717..289S}
Steidel C.~C.,  Erb D.~K.,  Shapley A.~E.,  Pettini M.,  Reddy N.,
  Bogosavljevi{\'c} M.,  Rudie G.~C.,    Rakic O.,  2010, The Astrophysical
  Journal, 717, 289

\bibitem[\protect\citeauthoryear{Stewart, Kaufmann, Bullock, Barton, Maller,
  Diemand \& Wadsley}{Stewart et~al.}{2011}]{2011ApJ...735L...1S}
Stewart K.~R.,  Kaufmann T.,  Bullock J.~S.,  Barton E.~J.,  Maller A.~H.,
  Diemand J.,    Wadsley J.,  2011, The Astrophysical Journal Letters, 735, L1

\bibitem[\protect\citeauthoryear{Stinson, Brook, Prochaska, Hennawi, Pontzen,
  Shen, Wadsley, Couchman, Quinn, Macci{\`o} \& Gibson}{Stinson
  et~al.}{2011}]{Stinson:2011tp}
Stinson G.,  Brook C.,  Prochaska J.~X.,  Hennawi J.,  Pontzen A.,  Shen S.,
  Wadsley J.,  Couchman H.,  Quinn T.,  Macci{\`o} A.~V.,    Gibson B.~K.,
  2011, arXiv, astro-ph.CO

\bibitem[\protect\citeauthoryear{Stinson, Seth, Katz, Wadsley, Governato \&
  Quinn}{Stinson et~al.}{2006}]{Stinson:2006id}
Stinson G.,  Seth A.,  Katz N.,  Wadsley J.,  Governato F.,    Quinn T.,  2006,
  Monthly Notices of the Royal Astronomical Society, 373, 1074

\bibitem[\protect\citeauthoryear{Stone \& Norman}{Stone \&
  Norman}{1992}]{Stone:1992iv}
Stone J.~M.,  Norman M.~L.,  1992, The Astrophysical Journal Supplement, 80,
  753

\bibitem[\protect\citeauthoryear{Tasker \& Bryan}{Tasker \&
  Bryan}{2006}]{Tasker:2006kp}
Tasker E.~J.,  Bryan G.~L.,  2006, The Astrophysical Journal, 641, 878

\bibitem[\protect\citeauthoryear{Tescari, Viel, D'Odorico, Cristiani, Calura,
  Borgani \& Tornatore}{Tescari et~al.}{2011}]{2011MNRAS.411..826T}
Tescari E.,  Viel M.,  D'Odorico V.,  Cristiani S.,  Calura F.,  Borgani S.,
  Tornatore L.,  2011, Monthly Notices of the Royal Astronomical Society, 411,
  826

\bibitem[\protect\citeauthoryear{Thacker \& Couchman}{Thacker \&
  Couchman}{2000}]{2000ApJ...545..728T}
Thacker R.~J.,  Couchman H. M.~P.,  2000, The Astrophysical Journal, 545, 728

\bibitem[\protect\citeauthoryear{Thom \& Chen}{Thom \&
  Chen}{2008}]{2008ApJ...683...22T}
Thom C.,  Chen H.-W.,  2008, The Astrophysical Journal, 683, 22

\bibitem[\protect\citeauthoryear{Thom, Werk, Tumlinson, Prochaska, Meiring,
  Tripp \& Sembach}{Thom et~al.}{2011}]{2011ApJ...736....1T}
Thom C.,  Werk J.~K.,  Tumlinson J.,  Prochaska J.~X.,  Meiring J.~D.,  Tripp
  T.~M.,    Sembach K.~R.,  2011, The Astrophysical Journal, 736, 1

\bibitem[\protect\citeauthoryear{Tripp, Sembach, Bowen, Savage, Jenkins, Lehner
  \& Richter}{Tripp et~al.}{2008}]{2008ApJS..177...39T}
Tripp T.~M.,  Sembach K.~R.,  Bowen D.~V.,  Savage B.~D.,  Jenkins E.~B.,
  Lehner N.,    Richter P.,  2008, ASTROPHYS J SUPPL S, 177, 39

\bibitem[\protect\citeauthoryear{Truelove, Klein, McKee, Holliman, Howell \&
  Greenough}{Truelove et~al.}{1997}]{Truelove:1997bj}
Truelove J.~K.,  Klein R.~I.,  McKee C.~F.,  Holliman J.~H.,  Howell L.~H.,
  Greenough J.~A.,  1997, The Astrophysical Journal Letters, 489, L179

\bibitem[\protect\citeauthoryear{Tumlinson, Thom, Werk, Prochaska, Tripp,
  Weinberg, Peeples, O'Meara, Oppenheimer, Meiring, Katz, Dav{\'e}, Ford \&
  Sembach}{Tumlinson et~al.}{2011}]{Tumlinson:2011wm}
Tumlinson J.,  Thom C.,  Werk J.~K.,  Prochaska J.~X.,  Tripp T.~M.,  Weinberg
  D.~H.,  Peeples M.~S.,  O'Meara J.~M.,  Oppenheimer B.~D.,  Meiring J.~D.,
  Katz N.~S.,  Dav{\'e} R.,  Ford A.~B.,    Sembach K.~R.,  2011, arXiv,
  astro-ph.CO

\bibitem[\protect\citeauthoryear{Turk, Smith, Oishi, Skory, Skillman, Abel \&
  Norman}{Turk et~al.}{2011}]{Turk:2011dd}
Turk M.~J.,  Smith B.~D.,  Oishi J.~S.,  Skory S.,  Skillman S.~W.,  Abel T.,
   Norman M.~L.,  2011, The Astrophysical Journal Supplement, 192, 9

\bibitem[\protect\citeauthoryear{Van De~Voort \& Schaye}{Van De~Voort \&
  Schaye}{2011}]{VanDeVoort:2011wk}
Van De~Voort F.,  Schaye J.,  2011, eprint arXiv:1111.5039

\bibitem[\protect\citeauthoryear{Van De~Voort, Schaye, Booth, Haas \&
  Dalla~Vecchia}{Van De~Voort et~al.}{2011}]{2011MNRAS.414.2458V}
Van De~Voort F.,  Schaye J.,  Booth C.~M.,  Haas M.~R.,    Dalla~Vecchia C.,
  2011, Monthly Notices of the Royal Astronomical Society, 414, 2458

\bibitem[\protect\citeauthoryear{Vazza, Br{\"u}ggen, Gheller \& Brunetti}{Vazza
  et~al.}{2012}]{2012MNRAS.421.3375V}
Vazza F.,  Br{\"u}ggen M.,  Gheller C.,    Brunetti G.,  2012, Monthly Notices
  of the Royal Astronomical Society, 421, 3375

\bibitem[\protect\citeauthoryear{Verner, Verner \& Ferland}{Verner
  et~al.}{1996}]{1996ADNDT..64....1V}
Verner D.~A.,  Verner E.~M.,    Ferland G.~J.,  1996, Atomic Data and Nuclear
  Data Tables, 64, 1

\bibitem[\protect\citeauthoryear{Wise, Abel, Turk, Norman \& Smith}{Wise
  et~al.}{2012}]{2012MNRAS.427..311W}
Wise J.~H.,  Abel T.,  Turk M.~J.,  Norman M.~L.,    Smith B.~D.,  2012,
  Monthly Notices of the Royal Astronomical Society, 427, 311

\bibitem[\protect\citeauthoryear{Yao, Nowak, Wang, Schulz \& Canizares}{Yao
  et~al.}{2008}]{2008ApJ...672L..21Y}
Yao Y.,  Nowak M.~A.,  Wang Q.~D.,  Schulz N.~S.,    Canizares C.~R.,  2008,
  The Astrophysical Journal, 672, L21

\end{thebibliography}

\label{lastpage}
\end{document}